\documentclass[aps,prb,longbibliography,twocolumn,floatfix,superscriptaddress,nofootinbib,10pt]{revtex4-2}

\usepackage{graphicx}
\usepackage{amsmath}
\usepackage{amssymb}
\usepackage{amsfonts}
\usepackage{mathtools}
\usepackage{physics}
\usepackage{xcolor} % for RGB colors
\usepackage[colorlinks=true, linkcolor=blue, citecolor=blue, urlcolor=blue]{hyperref}
\usepackage{booktabs}
\usepackage{tabularx}   % equal-width columns
\usepackage{booktabs}   % nicer rules (\toprule, \midrule, \bottomrule)
\usepackage{array} 
\usepackage{pgffor}    % for centering column types
\usepackage{comment}

\begin{document}
	
\title{Controlled Zeno-Induced Localization of Free Fermions in a Quasiperiodic Chain}

\author{Pinaki Singha}
\affiliation{Department of Physics, Indian Institute of Science Education and Research, Bhopal, Madhya Pradesh 462066, India}
\author{Nilanjan Roy}
\affiliation{Division of Physics and Applied Physics, Nanyang Technological University, Singapore 637371}
\affiliation{School of Physical Sciences, National Institute of Science Education and Research, Jatni 752050, India}
\affiliation{Homi Bhabha National Institute, Training School Complex, Anushaktinagar, Mumbai 400094, India}
\author{Marcin Szyniszewski}
\affiliation{Department of Computer Science, University of Oxford, Parks Road, Oxford OX1 3QD, UK}
\affiliation{Department of Physics and Astronomy, University College London, Gower Street, London, WC1E 6BT, UK}
\author{Auditya Sharma}
% \email{auditya@iiserb.ac.in}
\affiliation{Department of Physics, Indian Institute of Science Education and Research, Bhopal, Madhya Pradesh 462066, India}
\date{\today}
\begin{abstract}
    We investigate measurement-induced localization in a continuously monitored one-dimensional Aubry--Andr\'e--Harper model, focusing on the quantum Zeno regime in which the measurements dominate coherent dynamics. The presence of a quasiperiodic potential renders the problem analytically tractable and enables a controlled study of the interplay between monitoring and disorder.
    We develop an analytical description based on an instantaneous Schr\"odinger equation with a measurement-induced effective potential constructed self-consistently from individual quantum trajectories, without relying on postselection. In the quantum Zeno regime, an emergent dominant energy scale reduces the problem to a transfer-matrix formulation of an effective non-Hermitian Hamiltonian, which allows direct computation of the Lyapunov exponent.
    Complementarily, we extract the localization length numerically from long-time steady-state quantum state diffusion trajectories by reconstructing the intrinsic localized single-particle wave functions and analyzing their spatial decay.
    These numerical results show quantitative agreement with the effective theory predictions, with controlled corrections of order $J^2/[\lambda^2+(\gamma/2)^2]$ (where $J$ is the hopping amplitude, $\gamma$ the measurement strength, and $\lambda$ the quasiperiodic potential).
    Our results underscore the connection between the effective non-Hermitian description and the stochastic monitored dynamics, showing the interplay between Zeno-like localization, coherent hopping, and quasiperiodic-disorder-induced localization, while also laying the groundwork for understanding and exploiting measurement-induced localization as a tool for quantum control and state preparation.
\end{abstract}

\maketitle
\section{Introduction}
\label{sec:intro}

The dynamics of quantum many-body systems under continuous monitoring has emerged as a central theme in nonequilibrium quantum physics. Unlike closed systems, where unitary evolution governs transport and entanglement growth~\cite{Deutsch1991, Srednicki1994, DAlessio2016, Borgonovi2016}, monitored quantum systems exhibit qualitatively new behavior arising from the competition between coherent dynamics and measurement backaction. This competition gives rise to measurement-induced phase transitions, most notably the transition between volume-law and area-law entangled phases in hybrid quantum circuits~\cite{Chan2019, PhysRevX.9.031009, PhysRevB.98.205136, PhysRevB.100.134306, PhysRevB.100.064204, PhysRevLett.125.030505, 10.21468/SciPostPhysCore.5.2.023, PhysRevX.12.041002, PRXQuantum.4.040332, PhysRevLett.129.120604, 10.21468/SciPostPhys.15.6.250}. Related phenomena have also been extensively explored in condensed-matter systems, where continuous monitoring modifies transport, localization, and entanglement dynamics in both interacting and noninteracting fermionic lattice models~\cite{PhysRevResearch.7.023082, PhysRevResearch.7.023082, 10.21468/SciPostPhys.7.2.024, PhysRevResearch.2.033017, PhysRevB.103.174303, PhysRevB.105.094303, PhysRevResearch.4.033001, 10.21468/SciPostPhys.14.5.138, PhysRevB.108.L020306, PhysRevX.13.041046, PhysRevLett.126.170602, PhysRevB.106.L220304, PhysRevResearch.5.033174, PhysRevX.11.041004, PhysRevLett.126.123604, PhysRevB.106.024304, 10.21468/SciPostPhys.14.3.031, PhysRevX.13.041045, PhysRevB.105.064305, 10.21468/SciPostPhysCore.6.4.078, PhysRevB.108.104313, Turkeshi2021}. 

A particularly important regime of monitored dynamics is the strong-measurement (quantum Zeno) limit, in which frequent local measurements inhibit coherent hopping and effectively restrict the system's evolution to a reduced subspace of Hilbert space~\cite{PhysRevResearch.2.033512,Slichter_2016,PhysRevLett.89.080401}. This phenomenon, known as the quantum Zeno effect~\cite{Misra1977Zeno, PhysRevA.41.2295}, has been widely explored in contexts ranging from decoherence control to constrained quantum dynamics and measurement-stabilized phases~\cite{PhysRevB.98.205136,Biella2021manybodyquantumzeno}. From a dynamical perspective, the Zeno regime exhibits a strong suppression of transport~\cite{PhysRevLett.97.260402, PhysRevA.86.032120} and the emergence of localization-like behavior, even in systems that are otherwise delocalized~\cite{PhysRevLett.115.140402}. Beyond their fundamental interest, such Zeno-dominated dynamics are directly relevant to modern experimental platforms, including superconducting qubits with dispersive readout~\cite{Koh2023, PhysRevLett.117.190503}, trapped-ion systems based on fluorescence detection~\cite{Noel2022}, and ultracold atomic gases probed by quantum gas microscopes with single-site resolution~\cite{PhysRevLett.115.140402}. In these settings, measurement backaction is not only a source of decoherence but can be exploited as a dynamical resource to suppress transport, stabilize quantum states~\cite{Kondo2016}, and engineer effective constraints on the system's evolution, making the quantum Zeno regime a powerful framework for understanding measurement-induced localization~\cite{PhysRevLett.115.140402} and for designing controlled nonequilibrium quantum states in contemporary quantum simulation platforms~\cite{Slichter_2016,Koh2023}.

In parallel, localization phenomena in low-dimensional quantum systems have long served as a paradigm for understanding the suppression of transport due to random disorder~\cite{Anderson1958, Thouless1972, Abrahams1979, Basko2006, Gornyi2005, Pal2010, Nandkishore2015, Abanin2019} or quasiperiodicity~\cite{AubryAndre1980, Grempel1982, Thouless1983, Iyer2013}. The Aubry--Andr\'e--Harper (AAH) model, in particular, provides a minimal and experimentally relevant platform exhibiting a localization transition driven by quasiperiodic modulation~\cite{AubryAndre1980, g7vd-hgw4, Roati2008, Lahini2009}. While localization in closed systems has been extensively investigated, far less is known about how continuous monitoring modifies localization properties and transport in such systems~\cite{Lunt2020, Boorman2022, Tang2025, PhysRevB.108.165126, PhysRevB.110.024303}. In one-dimensional noninteracting fermionic systems with particle number conservation, measurement-induced entanglement transitions generically do not survive in the thermodynamic limit, leading instead to a localized area-law phase~\cite{10.21468/SciPostPhys.7.2.024, Fidkowski2021, PhysRevX.13.041046, PhysRevResearch.6.043246, jppz-vdgn, Poboiko2024, Poboiko2025}. Thus, monitored free fermions provide an ideal platform in which to address the open and timely question of how measurement backaction reshapes localization length scales and spatial structure in the quantum Zeno regime.

\begin{figure}[tbp]
	\centering
	\includegraphics[width=0.9\columnwidth]{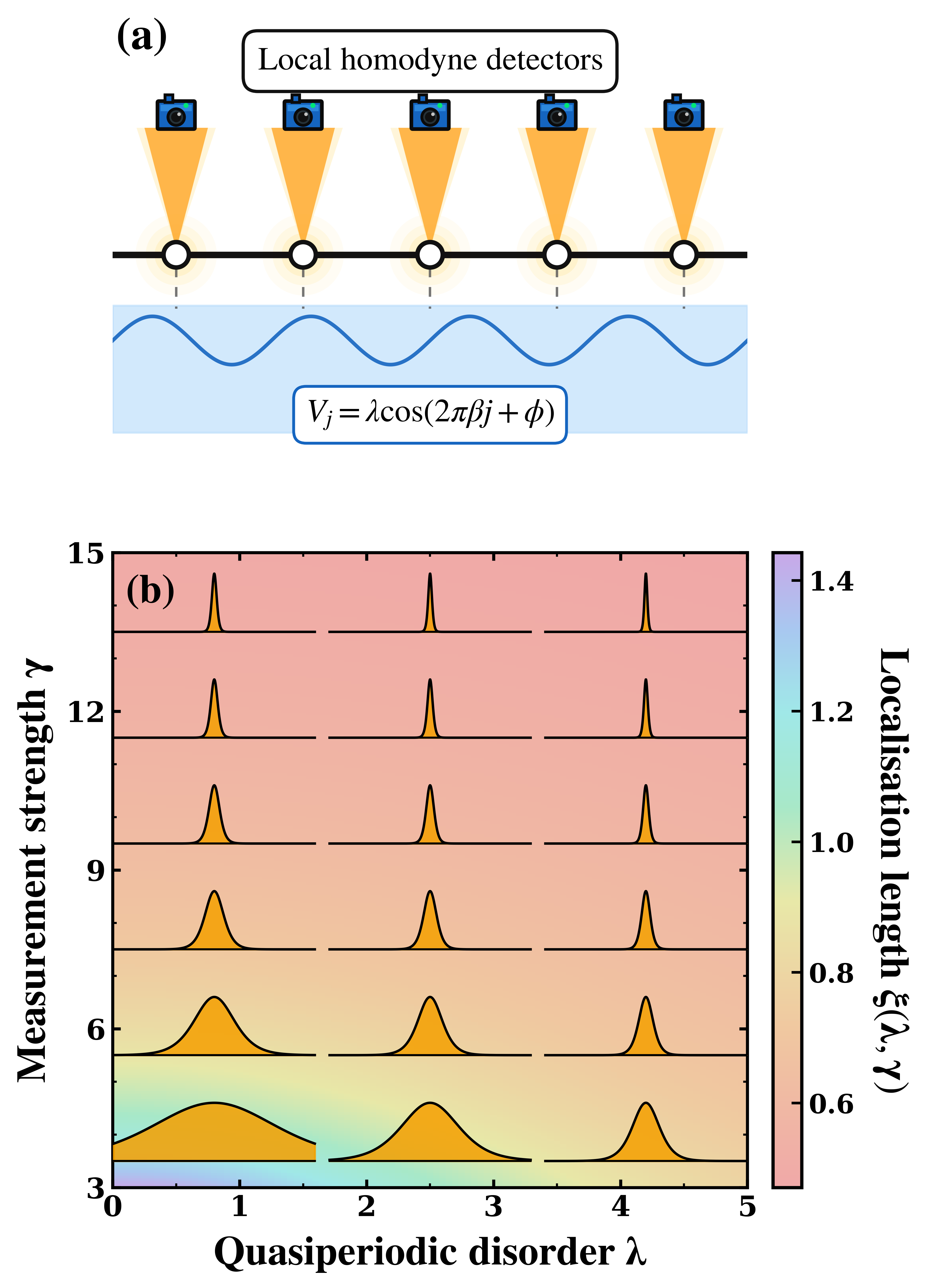}
	\caption{
		(a) Schematic of the Aubry--Andr\'e--Harper lattice under continuous homodyne detection.
		(b) Illustrative plot of how the interplay between the quasiperiodic disorder $\lambda$ and the measurement strength $\gamma$ localizes the single-particle wave function profiles (in orange). In the background, we also show the density plot of the localization length $\xi(\lambda, \gamma)$ extracted using an effective non-Hermitian theory developed in this work. Hopping amplitude is set to $J=1$.
        %Phase diagram of $\kappa(\lambda,\gamma)$ showing localization
%arising from the interplay between quasiperiodic disorder $\lambda$
%and measurement strength $\gamma$. Increasing either parameter enhances
%localization, as seen from the growth of $\kappa$ and the confined
%wavefunctions (insets).
	}
	\label{fig:combined}
\end{figure}

In this work, we study measurement-induced localization in a continuously monitored one-dimensional Aubry--Andr\'e--Harper model~\cite{3zfd-3hqt} [see Fig.~\ref{fig:combined}(a)] in the quantum Zeno regime using quantum state diffusion (QSD)~\cite{Percival1998, gisin1997quantumstatediffusionfoundations, Percival_1999}. By combining QSD simulations with an effective non-Hermitian description, we provide a controlled and quantitative characterization of the localization length, going beyond entanglement -- or transport-based approaches. Within the QSD framework, continuous measurement is described in terms of stochastic pure-state trajectories~\cite{Percival_1999}, enabling direct access to both dynamical and steady-state properties. We focus on how in the strong-measurement regime quasiperiodic modulation and measurement backaction shape the spatial structure of the wavefunction [see Fig.~\ref{fig:combined}(b)].

Rather than assuming selection of a single eigenmode, we find that long-time behavior is controlled by an emergent dominant energy scale that, in the Zeno limit, produces a clear separation of time scales so that non-dominant components relax rapidly and the dynamics admits an effective stationary description. Within the QSD picture, this can be interpreted as a measurement-induced modification of the effective potential that combines quasiperiodic lattice modulation and measurement backaction, reshaping the energy landscape and constraining the steady-state spatial structure of the wavefunction. Exploiting this stationary description, we derive a non-Hermitian Schr\"odinger equation, which (in contrast to other non-Hermitian effective descriptions that emerge from postselected dynamics~\cite{Turkeshi2021, 10.21468/SciPostPhys.14.3.031, 10.21468/SciPostPhys.14.5.138, DeTomasi2024}) requires no postselection and provides a controlled characterization of the order of corrections to the full QSD evolution.
This leads naturally to a transfer-matrix treatment of the monitored dynamics, showing that localization in the quantum Zeno regime [see Fig.~\ref{fig:combined}(b)] arises from the interplay of coherent hopping, quasiperiodicity, and measurement-induced dissipation rather than from a disorder-driven critical point. The resulting phase diagram reveals a Zeno-like regime where the time-scale separation holds, and our analytical theory applies (with localization lengths in excellent quantitative agreement with QSD trajectory numerics), and a weak-measurement regime where the separation is absent and QSD-induced localization is not directly relevant. Our effective theory further yields closed-form expressions for the localization length in the Zeno limit for both weak and strong disorder, providing a compact, quantitatively accurate description that reveals the underlying physics.

The remainder of the paper is organized as follows. Sec.~\ref{sec:model} introduces the model and the quantum state diffusion formalism, as well as the methodology for extracting the localization length numerically. Sec.~\ref{sec:qsd_heff_equivalence} derives the localization length in the quantum Zeno limit using the effective transfer-matrix description. Sec.~\ref{sec:results} compares the effective theory predictions with the QSD numerical results, and Sec.~\ref{sec:conclusion} summarizes our findings and discusses future directions.

\section{Model and methodology}
\label{sec:model}

\subsection{The Aubry--Andr\'e--Harper model}

We consider a one-dimensional chain of spinless fermions with open boundary conditions, described by the Aubry--Andr\'e--Harper (AAH) model~\cite{AubryAndre1980, g7vd-hgw4}; see Fig.~\ref{fig:combined}(a). The Hamiltonian is
\begin{equation}
    H_{\mathrm{AAH}} =
    - J \sum_{j=1}^{L-1} \left( c_j^\dagger c_{j+1} + \mathrm{h.c.} \right)
    + \sum_{j=1}^{L} V_j\, c_j^\dagger c_j ,
\end{equation}
where \(c_j^\dagger\) (\(c_j\)) creates (annihilates) a fermion on site \(j\), \(J\) is the nearest-neighbour hopping amplitude, and  \(V_j\) denotes a site-dependent onsite potential, which takes the quasiperiodic form
\begin{equation}
    V_j = \lambda \cos(2\pi \beta j + \phi),
\end{equation}
where \(\lambda\) is the potential strength, \(\beta\) is an irrational modulation wavevector (typically chosen as the inverse golden ratio to ensure quasiperiodicity), and \(\phi\) is a phase offset. The AAH model exhibits a localization transition at \(\lambda = 2J\), separating extended single-particle eigenstates from exponentially localized ones. Unlike random-disorder-driven localization, the AAH model phase transition is controlled by a deterministic incommensurate potential, making it analytically tractable and convenient for studies of Anderson localization transition in disordered systems.

\subsection{Quantum state diffusion}

To study the effects of continuous local measurements, we employ the quantum state diffusion formalism, which provides a stochastic unraveling of the Lindblad master equation in terms of pure-state trajectories~\cite{PhysRevA.69.032107,GisinPercival1992,Percival1998}. The time evolution of the state \(|\psi(t)\rangle\) is governed by the stochastic Schr\"odinger (SSE) equation
\begin{align}
    d|\psi(t)\rangle &= 
    \Bigg[
    - i H_{\mathrm{AAH}}
    - \frac{\gamma}{2} \sum_{j=1}^{L} \left( n_j - \langle n_j \rangle \right)^2
    \Bigg] dt \, |\psi(t)\rangle  \nonumber\\
    & \quad + \sum_{j=1}^{L} \sqrt{\gamma}\,
    \left( n_j - \langle n_j \rangle \right)
    \, dW_j(t)\, |\psi(t)\rangle ,
    \label{eq:aah_qsd}
\end{align}
where \(n_j = c_j^\dagger c_j\) is the local density operator and \(\langle n_j \rangle = \langle \psi(t) | n_j | \psi(t) \rangle\) is its expectation value. The stochastic increments \(dW_j(t)\) are independent complex Wiener processes satisfying \(\mathbb{E}[dW_j(t)] = 0\) and \(\mathbb{E}[dW_j(t)\, dW_k^*(t)] = \delta_{jk}\, dt\). 
This SSE with local density measurements can be viewed as continuously monitoring the system with homodyne detectors, as illustrated in Fig.~\ref{fig:combined}(a).
Within the QSD framework, the coherent evolution generated by \(H_{\mathrm{AAH}}\) is supplemented by stochastic measurement backaction associated with continuous monitoring of the local density operators \(n_j = c_j^\dagger c_j\), with measurement strength \(\gamma\). Individual quantum trajectories thus encode both the deterministic drift and stochastic fluctuations induced by the measurement process, while averaging over trajectories reproduces the ensemble Lindblad dynamics.
%Averaging over an ensemble of such trajectories reproduces the Lindblad dynamics, while individual trajectories encode the measurement backaction at the level of pure states.
Note that observables that reveal measurement-induced transitions involve nonlinear functions of the density matrix, and are therefore accessible within QSD, but not within the Lindblad averaged dynamics.

%\subsection{Numerical Implementation}

The monitored dynamics considered in this work conserve particle number and can be efficiently simulated within the SSE framework. At all times, the many-body wavefunction remains a pure Gaussian state of \(N\) fermions on \(L\) lattice sites and can be represented as~\cite{10.21468/SciPostPhys.7.2.024, PhysRevLett.126.170602}:
\begin{equation}
	|\psi(t)\rangle
	=
	\prod_{k=1}^{N}
	\left(
	\sum_{j=1}^{L} U_{jk}(t)\, c_j^\dagger
	\right)
	|0\rangle ,
	\label{eq:slater}
\end{equation}
where \(U(t)\) is an \(L \times N\) matrix whose columns correspond to orthonormal single-particle wave functions (Slater determinant matrix, orbital matrix)~\cite{PhysRevB.108.165126, PhysRevB.110.024303}, \(c_j^\dagger\) are fermionic creation operators, and \(|0\rangle\) denotes the vacuum. Throughout this work we focus on the half-filled case and initialize the evolution from a N\'eel state. All single-particle observables are efficiently obtained from the equal-time correlation matrix
\begin{equation}
    D(t) = U(t) U^\dagger(t) .
    \label{eq:correlation_matrix}
\end{equation}
For numerical implementation, the SSE evolution over a small time step \(dt\) is approximated using a Trotter decomposition
\begin{equation}
	|\psi(t+dt)\rangle \simeq e^{\mathcal{M}} e^{-i H_{\mathrm{AAH}} dt} |\psi(t)\rangle .
	\label{eq:trotter_manybody}
	\end{equation}
	Owing to the Gaussian structure of the state, this evolution translates directly into an update of the orbital matrix,
	\begin{equation}
	U(t+dt) = e^{M} e^{-i h_{\mathrm{AAH}} dt} U(t) ,
	\label{eq:orbital_update}
\end{equation}
where \(h_{\mathrm{AAH}}\) denotes the single-particle representation of \(H_{\mathrm{AAH}}\). The measurement backaction is encoded in the diagonal matrix \(M\), whose elements are given by~\cite{PhysRevB.108.165126}
\begin{equation}
	M_{ij}
	=
	\delta_{ij}
	\left[
	\eta_j
	+
	\gamma \big( 2 \langle n_j \rangle_t - 1 \big) dt
	\right],
	\label{eq:measurement_matrix}
\end{equation}
with \(\eta_j\) real Gaussian noise variables drawn from a normal distribution of variance \(\gamma dt\). After each time step, the updated orbital matrix is reorthonormalized via a QR decomposition, ensuring numerical stability and preserving the Gaussian character of the state throughout the stochastic evolution.

\subsection{Localization length extraction}
\label{subsec:loc_length}

In this subsection, we describe how localization lengths are extracted from steady-state QSD trajectories. This procedure provides the numerical benchmark against which we will later test the analytical effective theory developed in Sec.~\ref{sec:qsd_heff_equivalence}. Because the QSD evolution mixes the orbital basis through orthonormalization, a careful unscrambling procedure is required before physically meaningful localization properties can be obtained.

In particle-number-conserving free-fermion systems, the many-body wavefunction can be written as a Slater determinant of single-particle orbitals~\cite{PhysRevB.108.165126,PhysRevB.110.024303}. This representation is not unique: although the single-particle correlation matrix fully determines all physical observables, the choice of the orbital basis itself is defined only up to arbitrary unitary rotations within the occupied subspace.

In numerical simulations based on QSD, the time evolution is nonunitary due to measurement backaction. To ensure numerical stability, the evolving orbitals (specifically, the matrix $U$) must therefore be reorthonormalized at each time step. In our implementation, this is achieved using a QR decomposition. While this procedure preserves orthonormality, it also introduces additional unitary rotations among the occupied orbitals, leading to an arbitrary mixing of the Slater-determinant orbitals during the evolution.
% NOTE: Not really needed, since this is already in the intro of Ref. PhysRevB.110.024303.
% Mathematically, the occupied orbitals are collected in a matrix \(U \in \mathbb{C}^{L \times N}\), where \(U_{in}\) denotes the amplitude of orbital \(n\) on lattice site \(i\). The non-uniqueness of the Slater-determinant representation is expressed by the transformation
% \begin{equation}
% U \;\longrightarrow\; U' = U Q, \qquad Q^\dagger Q = \mathbb{I},
% \end{equation}
% where \(Q \in \mathbb{C}^{N \times N}\) is an arbitrary unitary matrix acting in orbital space. Such transformations mix the orbitals among themselves but leave the many-body state and all physical observables unchanged. This invariance follows from the fact that observables depend only on the single-particle correlation matrix $D$, which is unchanged under orbital rotations,
% \begin{equation}
% D' = U' U'^\dagger = U Q Q^\dagger U^\dagger = D.
% \end{equation}
Although this does not affect the physical state, it obscures the intrinsic spatial structure of the individual orbitals. As a result, orbitals obtained directly from the QSD evolution may appear artificially delocalized, even when the underlying many-body state is localized. To access physically meaningful localization properties, we therefore fix this orbital gauge freedom by applying the orbital unscrambling procedure of Ref.~\cite{PhysRevB.110.024303} to steady-state QSD trajectories, yielding a maximally localized orbital basis, which provides a physically meaningful description of the system, e.g.\@ through orbital shape or inverse participation ratio.

Once a localized orbital basis is obtained, we extract the localization length $\xi$ by analyzing the spatial decay of individual orbitals. Each orbital corresponds to a column of the matrix \(U\), with \(i=1,\dots,L\) labeling lattice sites and \(n=1,\dots,N\) labeling orbitals. For a given orbital \(n\), we identify the site at which its amplitude is maximal, $x_{\max}^{(n)}$, %
%\begin{equation}
%    x_{\max}^{(n)} = \arg\max_i |U_{in}|,
%\end{equation}
which defines the localization center. Spatial decay is then analyzed as a function of the distance $k = |i - x_{\max}^{(n)}|$ from this center. %To this end, we introduce a nonnegative integer distance
%\begin{equation}
%    k = |i - x_{\max}^{(n)}|,
%\end{equation}
%which counts the number of lattice sites away from the localization center.

Under open boundary conditions, the decay profile can differ on the two sides of the maximum. We therefore treat the right and left sides separately. The right-tail amplitudes are defined as
\begin{equation}
    U_n^{\mathrm{right}}(k) = |U_{(x_{\max}^{(n)} + k)n}|, 
    \qquad k = 0,1,\dots,L-1-x_{\max}^{(n)},
\end{equation}
while the left-tail amplitudes are
\begin{equation}
    U_n^{\mathrm{left}}(k) = |U_{(x_{\max}^{(n)} - k)n}|, 
    \qquad k = 0,1,\dots,x_{\max}^{(n)}.
\end{equation}
For each tail, we analyze the spatial decay by first averaging the logarithm of the orbital amplitudes over localized orbitals at a fixed distance \(k\) from the localization center. The resulting quantity is then averaged over independent stochastic trajectories. The doubly averaged profiles are fitted to the linear forms
\begin{equation}
    \label{eq:log_decay_fits}
    \begin{aligned}
    \bigl\langle \langle \ln |U_n^{\mathrm{right}}(k)| \rangle_{\mathrm{orb}} \bigr\rangle_{\mathrm{traj}}
    &= m_{\mathrm{right}}\, k + c_{\mathrm{right}}, \\
    \bigl\langle \langle \ln |U_n^{\mathrm{left}}(k)| \rangle_{\mathrm{orb}} \bigr\rangle_{\mathrm{traj}}
    &= -m_{\mathrm{left}}\, k + c_{\mathrm{left}} .
    \end{aligned}
\end{equation}
From the fitted slopes, we define the decay lengths associated with the right and left tails as
\begin{equation}
    \xi_{\mathrm{right}} = -\frac{1}{m_{\mathrm{right}}}, 
    \qquad
    \xi_{\mathrm{left}} = \frac{1}{m_{\mathrm{left}}}.
\end{equation}
The localization length is then defined as the average of the two,
\begin{equation}
    \xi = \frac{1}{2}\left(\xi_{\mathrm{left}} + \xi_{\mathrm{right}}\right),
\end{equation}
which corresponds to the inverse Lyapunov exponent governing the exponential decay of localized orbitals.

Another key quantity of interest will be the Lyapunov exponent $\kappa$, which quantifies the rate at which nearby trajectories in phase space diverge under evolution. In lattice models or quasiperiodic systems such as the AAH model, it can also be defined for transfer matrices, where it characterizes the exponential decay of the wavefunction~\cite{Comtet2013} and is directly related to the localization length via
\begin{equation}
    \kappa = \xi^{-1}.
\end{equation}
	
The localization lengths obtained through this procedure serve as the quantitative reference for the analytical predictions derived in Sec.~\ref{sec:qsd_heff_equivalence}. In particular, the QSD-based values of $\xi$ and $\kappa$ will be compared directly with the corresponding quantities obtained from the effective non-Hermitian description, as shown in Fig.~\ref{fig:xi_four_panels} in Sec.~\ref{sec:results}. This establishes the link between the numerical steady-state dynamics and the theoretical framework developed in the following section.

	% Required packages (add to your preamble if not already present):
	% \usepackage{amsmath,amssymb}
	% \usepackage{graphicx}
	% \usepackage{subcaption}   % optional, for multi-panel figures
	% \usepackage{braket}       % optional, for \ket and \bra macros

%\section{Analytical Derivation of the Localization Length in the Zeno Limit}
\section{Effective non-Hermitian description of the monitored dynamics}
\label{sec:qsd_heff_equivalence}

We now develop an analytical description of the steady-state localization observed in the monitored AAH chain. The central idea is that, in the quantum Zeno regime, strong local measurements suppress coherent hopping and produce a separation of time scales: typical QSD trajectories concentrate near a sharply defined dominant energy while their spatial profiles remain localized up to controlled fluctuations. This allows us to construct an instantaneous effective description in which the monitored dynamics is represented by a non-Hermitian potential reconstructed from typical trajectory snapshots, without invoking postselection. We begin by showing that the long-time dynamics is governed by a dominant energy scale. Next, we describe and analyze fluctuations of a localized state within a manifold of pointer states that dominate the long-time dynamics in the quantum Zeno regime. This allows us to write an effective potential and its fluctuations emerging from the QSD evolution. Finally, using the transfer-matrix formalism, we determine the Lyapunov exponent and the corresponding localization length.

\subsection{Emergence of the dominant energy scale}
\label{subsec:dominant_energy}

The first step in reducing the stochastic monitored dynamics to an effective localization problem is to identify the energy scale about which typical long-time trajectories fluctuate. Since the QSD evolution is stochastic, the instantaneous energy along a trajectory is not conserved in the usual closed-system sense. Nevertheless, in the quantum Zeno regime, the monitored dynamics reaches a nonequilibrium steady state whose trajectory-resolved energy distribution becomes sharply peaked in the thermodynamic limit. Establishing this self-averaging property allows us to replace the time-dependent instantaneous energy by a dominant value $E_{\mathrm{dom}}$, which will serve as the reference energy in the effective Schr\"odinger description developed below.

%In the quantum Zeno regime, continuous monitoring drives the system into a nonequilibrium steady state with time-independent statistical properties, despite the intrinsically stochastic nature of the dynamics. The evolution is described within the QSD formalism, which provides a trajectory-level representation of the monitored dynamics. 
For a quadratic fermionic Hamiltonian ($H = \sum_{i,j} H_{ij}\, c_i^\dagger c_j$) and a Gaussian many-body state parametrized by the time-dependent correlation matrix $D(t)$, the instantaneous energy per particle along a single QSD trajectory is given by
\begin{equation}
    E_{\mathrm{inst}}(t)
    =
    \frac{1}{N}
    \!\left[\operatorname{Tr}\!\left(H D(t)\right)\right].
\end{equation}
This expression yields the exact energy expectation value associated with an individual stochastic realization of the continuously monitored evolution.

\begin{figure}[t]
	\centering
	\includegraphics[width=\columnwidth]{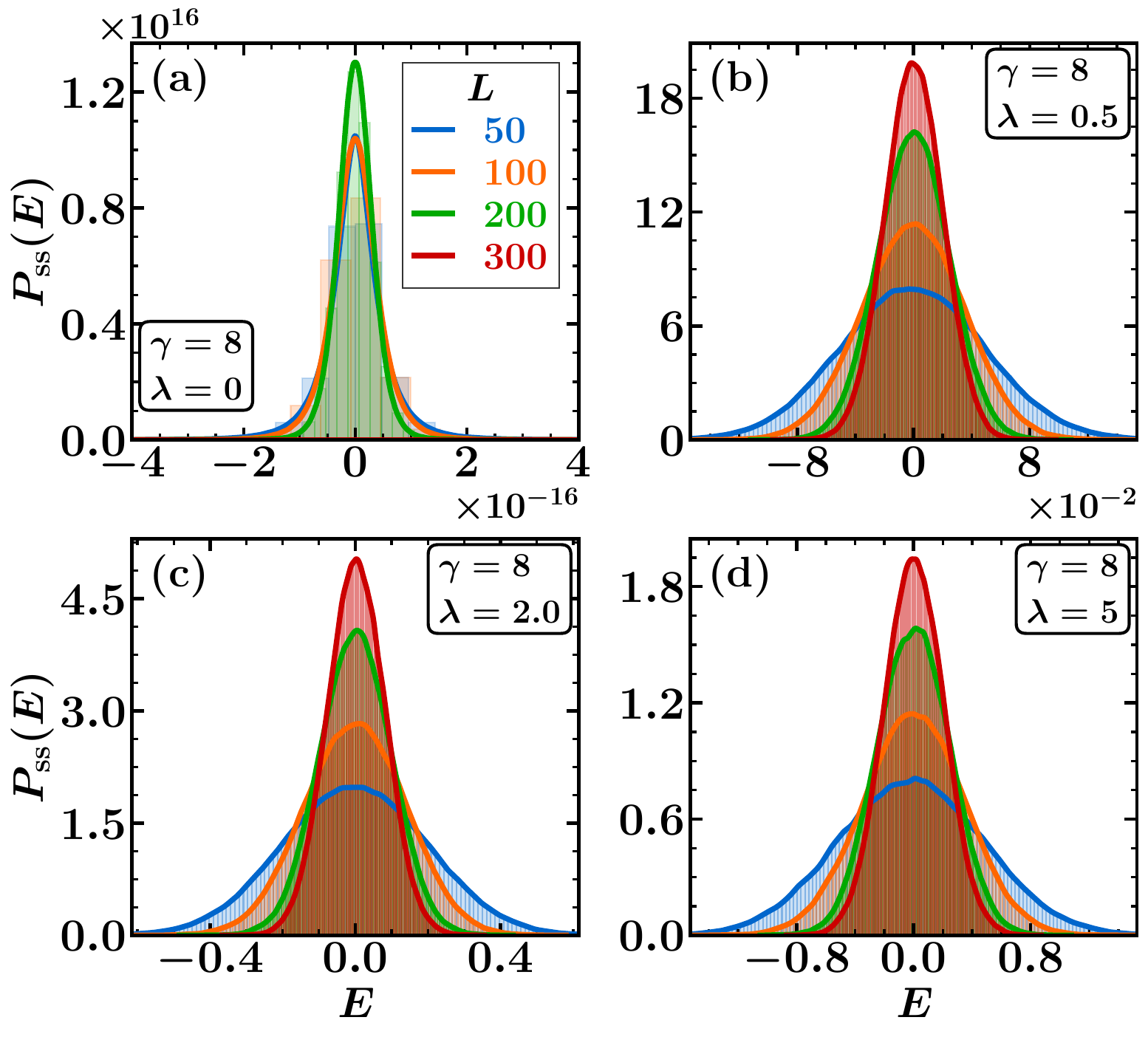}
	
	\vspace{0.02em}
	
	\includegraphics[width=0.9\columnwidth]{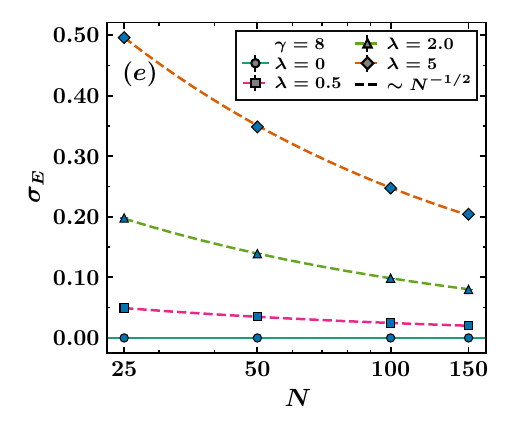}
	
    \caption{
    \textbf{Steady-state energy statistics under continuous monitoring}
    (\(\gamma = 8\), \(J = 1\)).
    \textbf{(Top)} Stationary distributions \(P_{\mathrm{ss}}(E)\) of the instantaneous energy for system sizes \(L = 50, 100, 200, 300\).
    Shaded regions show normalized steady-state histograms, and solid lines denote kernel density estimates.
    Panels correspond to the quasiperiodic potential strength of
    (a)~\(\lambda = 0\) (clean system),
    (b)~\(\lambda = 0.5\),
    (c)~\(\lambda = 2.0\),
    and (d)~\(\lambda = 5.0\).
    \textbf{(Bottom)}(e) Finite-size scaling of the energy fluctuations \(\sigma_E\) with particle number \(N\).
    Symbols indicate numerical results for different \(\lambda\), while the dashed line shows the theoretical scaling \(\sigma_E \propto N^{-1/2}\) [Eq.~\eqref{Fluc}].
    For \(\lambda = 0\), fluctuations vanish and are size independent, whereas for \(\lambda > 0\) the instantaneous energy is self-averaging in the thermodynamic limit.
    Error bars are smaller than the symbol size.
    }
    \label{fig:energy_statistics_gamma8}
\end{figure}

To characterize the long-time steady state, we define the stationary energy distribution by long-time averaging along a single QSD trajectory,
\begin{equation}
    P_{\mathrm{ss}}(E)
    =
    \lim_{T\to\infty}
    \frac{1}{T}
    \int_{t_0}^{t_0+T}
    \delta\!\bigl(E - E_{\mathrm{inst}}(t)\bigr)\,dt ,
\end{equation}
where $t_0$ denotes the equilibration time to reach the steady state, and $T$ is the total evolution time.
A central quantity extracted from \(P_{\mathrm{ss}}(E)\) is the \emph{dominant energy} \(E_{\mathrm{dom}}\), defined as the mode of the stationary energy distribution. Unlike the mean energy, which can be influenced by rare fluctuations, \(E_{\mathrm{dom}}\) characterizes the typical energetic sector most often visited by the monitored dynamics at long times and therefore provides a natural reference energy for an effective description.

%To characterize the long-time steady state, the QSD equations are integrated numerically up to a total evolution time \(T_{\mathrm{tot}}\), and the instantaneous energy \(E_{\mathrm{inst}}(t_n)\) is recorded at discrete time steps \(t_n = n\Delta t\).
%Transient effects from the initial state are removed by discarding an initial burn-in portion of the time series, after which the remaining data are interpreted as samples drawn uniformly in time from the stationary regime.
%and is estimated numerically using normalized histograms or kernel density estimation. When multiple independent QSD trajectories are simulated, the resulting steady-state samples are pooled together. Assuming ergodicity of the stationary QSD process for the observables considered, this procedure is equivalent to sampling the invariant measure of the stochastic dynamics.

Fig.~\ref{fig:energy_statistics_gamma8} shows $P_{\mathrm{ss}}(E)$ extracted numerically by discretizing the QSD evolution. In the clean system [panel (a)], translational invariance fixes the energy exactly, so $P_{\mathrm{ss}}(E)$ collapses to a sharp peak at $E=0$. For non-zero quasiperiodic potential strengths [panels (b)--(d)], finite-size fluctuations appear but shrink with increasing system size, and the distributions become progressively more sharply peaked at $E=0$, demonstrating self-averaging~\cite{PhysRevB.102.094310, Gordillo_Guerrero_2007, PhysRevB.101.174312}. We also find that the width $\sigma_E$ of the energy distribution scales as $N^{-1/2}$ [panel (e)], which is consistent with analytical calculations in Appendix~\ref{app:self_averaging}.

As a result, relative energy fluctuations vanish in the thermodynamic limit, and the stationary energy distribution converges, in the sense of weak convergence, to a delta function:
\begin{equation}
    P_{\mathrm{ss}}(E)\xrightarrow[L\to\infty]{}\delta\!\bigl(E-E_{\mathrm{dom}}\bigr).
\end{equation}
The emergence of a sharply defined dominant energy
is the first ingredient that makes a static effective description possible: the stochastic time dependence of the energy becomes irrelevant at leading order in the Zeno regime.
%has direct implications for the effective description of the monitored dynamics.
The stochastic, time-dependent energy \(E_{\mathrm{inst}}(t)\) entering the instantaneous Schr\"odinger equation may therefore be replaced by its dominant value \(E_{\mathrm{dom}}\). This replacement is exact in the clean system due to symmetry and is asymptotically justified for \(\lambda \neq 0\) by self-averaging in the thermodynamic limit. For the monitored AAH model, we find \(E_{\mathrm{dom}} \approx 0\), providing a reference energy for the subsequent analytical treatment.

\subsection{Fluctuations in the pointer space}

Having identified the dominant energy sector, we next characterize the residual stochastic motion of the QSD trajectory within the localized steady-state manifold. Strong continuous monitoring suppresses coherent hopping but does not dynamically select a unique single-particle orbital.
Instead, the monitored dynamics supports a degenerate manifold of equivalent localized pointer configurations~\cite{PhysRevResearch.2.033512}, which are explored by any QSD trajectory in the steady state. 
Our analysis, therefore, does not rely on identifying a dominant state, but rather on describing fluctuations within this pointer manifold.
The purpose of this subsection is to formulate a controlled expansion around an arbitrary localized reference state in this manifold and to derive the stochastic equations governing the orthogonal fluctuations. This step is essential as the effective potential constructed later is valid only if these fluctuations remain parametrically small in the Zeno regime.

Consider a normalized QSD trajectory parametrized by the instantaneous single-particle wavefunction $\psi(t)=\{\psi_j(t)\}_{j=1}^L$. We introduce $\varphi_0$ as an arbitrary localized reference configuration drawn from the pointer manifold~\cite{PhysRevX.8.021005}. $\psi(t)$ may then be decomposed as
\begin{equation}
    \psi(t) = \varphi_0 + \delta\psi(t),
    \label{eq:inst_wfn_decomp}
\end{equation}
where $\delta\psi(t)$ lies in the orthogonal complement of $\varphi_0$, $\langle \varphi_0 | \delta\psi(t) \rangle = 0$.
The fluctuation field $\delta\psi(t)$ may be expanded in an arbitrary orthonormal basis $\{\varphi_\alpha\}_{\alpha \ge 1}$ as
\begin{equation}
    \delta\psi(t) = \sum_{\alpha \ge 1} c_\alpha(t)\,\varphi_\alpha,
    \label{eq:fluct_field_expansion}
\end{equation}
where the fluctuation amplitudes are $c_\alpha(t)=\langle\varphi_\alpha|\delta\psi(t)\rangle$.
While such a decomposition is formally always possible, the Zeno regime ensures that the total fluctuation weight $\|\delta\psi(t)\|$ is parametrically small.
Our analytic treatment determines the scale governing this suppression and does not depend on the detailed structure of individual fluctuation components.

The reference state $\varphi_0$ is exponentially localized~\cite{PhysRevB.110.184211,jppz-vdgn} around a site $j_0$, with localization length $\xi = \mathcal{O}(1)$.
Within the localization core $|j-j_0|\lesssim\xi$, the amplitude satisfies $|\varphi_0(j)| = \mathcal{O}(1)$, while in the tails (see Appendix~\ref{app:localization_proof})
\begin{equation}
    |\varphi_0(j)|
    \le
    C\,\varepsilon\,e^{-|j-j_0|/\xi},
    \qquad
    \varepsilon \equiv
    \frac{J}{\sqrt{\lambda^2+(\gamma/2)^2}} \ll 1 ,
\end{equation}
where $C$ is a constant independent of system size. 
The small parameter $\varepsilon$ quantifies the suppression of coherent hopping by measurement and diagonal detuning. The fluctuation states $\{\varphi_\alpha\}_{\alpha\ge1}$ are also exponentially localized, with localization lengths of order $\xi$. 
Within the localization core of $\varphi_0$, their amplitudes satisfy $|\varphi_\alpha(j)| = \mathcal{O}(\varepsilon)$, ensuring that the fluctuation field $\delta\psi$ is parametrically small in the Zeno regime (see Appendix~\ref{app:localization_proof}).

We now aim to project the normalized QSD equation in Eq.~\eqref{eq:aah_qsd} onto the subspace orthogonal to a localized reference mode $\varphi_0$ and derive the effective stochastic dynamics of the corresponding fluctuation amplitudes.
% For local measurement operators $L_j=\sqrt{\gamma}\,n_j$, the normalized QSD
% equation in It\^o form reads
% \begin{align}
% d|\psi\rangle
% &=
% \Biggl[
% -iH
% -\frac{1}{2}\sum_j
% \bigl(L_j-\langle L_j\rangle\bigr)^2
% \Biggr]
% |\psi\rangle\,dt\nonumber
% \\
% &\quad
% +\sum_j
% \bigl(L_j-\langle L_j\rangle\bigr)
% |\psi\rangle\,dW_j(t),
% \label{eq:qsd_normalized}
% \end{align}
% \marcin{Is this just Eq.~\ref{eq:aah_qsd}?}
% with independent Wiener increments satisfying $\langle dW_j(t)\,dW_{j'}(t)\rangle=\delta_{jj'}\,dt$.
First, projecting Eq.~\eqref{eq:aah_qsd} onto the site basis $\psi_j(t)=\langle j|\psi(t)\rangle$ yields the exact component-wise It\^o stochastic equation
\begin{align}
    d\psi_j
    &=
    - i (h_{AAH}\psi)_j\,dt\nonumber\\
    &\quad-\frac{\gamma}{2}\psi_j\,dt
    + \gamma |\psi_j|^2 \psi_j\,dt
    -\frac{\gamma}{2}\sum_k |\psi_k|^4 \psi_j\,dt
    \nonumber\\
    &\quad
    + \sum_r \sqrt{\gamma}\,
    \bigl(\delta_{jr}\psi_r - |\psi_r|^2\psi_j\bigr)\,dW_r(t),
    \label{eq:qsd_component}
\end{align}
where $(h_{AAH}\psi)_j=\sum_k (H_{AAH})_{jk}\psi_k$. Next, we use the decomposition in Eq.~\eqref{eq:inst_wfn_decomp} and the expansion~\eqref{eq:fluct_field_expansion} of the fluctuation field in an orthonormal basis.
% We decompose the instantaneous state as in Eq.~\eqref{eq:inst_wfn_decomp}, and introduce the orthogonal projector
% \begin{equation}
% P = 1 - |\varphi_0\rangle\langle\varphi_0| .
% \end{equation}
% Expanding the fluctuation field in an orthonormal basis $\{\varphi_\alpha\}_{\alpha\ge1}$ of the projected subspace, $\delta\psi(t)=\sum_{\alpha\ge1}c_\alpha(t)\varphi_\alpha$, the fluctuation amplitudes are defined by
% \begin{equation}
% c_\alpha(t)=\langle\varphi_\alpha|\delta\psi(t)\rangle .
% \end{equation}
%Since the basis states are time independent, their evolution follows from the projection of the QSD increment,
%\begin{equation}
%dc_\alpha=\langle\varphi_\alpha|P\,d\psi\rangle ,
%\end{equation}
%where $P = 1 - |\varphi_0\rangle\langle\varphi_0|$ is the orthogonal projector.
Since the basis states $\{\varphi_\alpha\}_{\alpha\ge1}$ are time independent and orthogonal to the reference mode $\varphi_0$, the evolution of the fluctuation amplitudes is obtained by projection,
\begin{equation}
    dc_\alpha = \langle \varphi_\alpha | d\psi \rangle .
\end{equation}
Substituting Eq.~\eqref{eq:aah_qsd} and expanding all terms to linear order in the amplitudes $c_\alpha$, we obtain a closed \textit{stochastic equation for the fluctuations},
\begin{align}
    dc_\alpha
    &=
    -\sum_\beta M_{\alpha\beta} c_\beta\,dt
    + \sum_j \!\Big[
    \sum_\beta (A_j)_{\alpha\beta} c_\beta + u_{j,\alpha}
    \Big] dW_j(t)
    \nonumber\\
    &\quad
    + \mathcal{O}(\|c\|^2).
    \label{eq:lin_sde_c}
\end{align}
Eq.~\eqref{eq:lin_sde_c} follows from the orthogonal projection of the QSD equation onto the fluctuation subspace and linearization in the amplitudes $c_\alpha$. The drift matrix $M$ contains Hamiltonian, measurement-induced, and It\^{o}-correction contributions, while $A_j$ and $u_{j,\alpha}$ arise from the multiplicative and additive components of the measurement noise, respectively. Quadratic and higher-order terms in $c_\alpha$ are neglected: although $\mathbb{E}|c_\alpha|^2 = \mathcal{O}(1)$ in the stationary state, their contribution in real space is suppressed by the $\mathcal{O}(\varepsilon)$ amplitude of each fluctuation mode in the Zeno regime, yielding only $\mathcal{O}(\varepsilon^2)$ corrections, as shown explicitly in Sec.~\ref{subsec:asymptotic_fluctuations}. 
Thus, Eq.~\eqref{eq:lin_sde_c} bridges the microscopic QSD dynamics and the effective static theory: if its solutions remain controlled in the Zeno regime, then the reconstructed potential will differ from its leading static form only perturbatively.

\subsection{Asymptotic properties of the fluctuations}
\label{subsec:asymptotic_fluctuations}

We now estimate the size and stability of the fluctuation amplitudes introduced above. The goal is not to solve the full stochastic fluctuation dynamics exactly, but to determine how its contributions scale with the small parameter $\varepsilon = J/\sqrt{\lambda^2+(\gamma/2)^2}$. Specifically, we consider each term of the stochastic equation for the fluctuations in Eq.~\eqref{eq:lin_sde_c} and their asymptotic behavior in the Zeno limit. We find that the resulting fluctuation dynamics is stable, and the fluctuation field remains $\mathcal{O}(\epsilon)$, providing the central control parameter for the effective potential constructed in the next subsection.
%Let us now consider each term of the stochastic equation for the fluctuations~\eqref{eq:lin_sde_c} and their asymptotic behavior in the Zeno limit.

The stochastic terms in Eq.~\eqref{eq:lin_sde_c} separate naturally into additive ($u_j$) and multiplicative ($A_j$) contributions. Explicit evaluation of $\langle \varphi_\alpha | \sqrt{\gamma} (n_j-\langle n_j \rangle) | \psi \rangle$ to linear order in $c_\alpha$ yields the following expressions for $A_j$ and $u_j$:
\begin{align}
    u_{j,\alpha}
    &=
    \sqrt{\gamma}\,
    \varphi_\alpha^*(j)\,\varphi_0(j), \\[4pt]
    (A_j)_{\alpha\beta}
    &=
    \sqrt{\gamma}\,
    \bigl[
    \varphi_\alpha^*(j)\,\varphi_\beta(j)
    -\delta_{\alpha\beta}a_j
    \bigr],
    \label{eq:Aj}
\end{align}
with $a_j=|\varphi_0(j)|^2$. The vector $u_j$ represents an additive noise source originating from the overlap between the reference state and the fluctuation modes at site $j$, while the matrices $A_j$ encode multiplicative noise through their linear coupling to the fluctuation amplitudes. The subtraction of the diagonal term proportional to $a_j$ reflects the centering of the measurement operators and ensures norm preservation of the stochastic evolution.

The deterministic drift matrix $M$ in Eq.~\eqref{eq:lin_sde_c} receives contributions from the Hamiltonian, the projected measurement drift, and the It\^o correction generated by multiplicative noise. 
The measurement-induced deterministic term is obtained by projecting $-\tfrac{\gamma}{2}\sum_j(n_j-\langle n_j\rangle)^2$ from Eq.~\eqref{eq:aah_qsd} onto the fluctuation subspace, which yields
\begin{equation}
    \label{eq:Q_def_j}
    Q_{\alpha\beta}
    =
    -\frac{\gamma}{2}\sum_j
    \bigl[
    \varphi_\alpha^*(j)\,\varphi_\beta(j)
    -2a_j\varphi_\alpha^*(j)\,\varphi_\beta(j)
    +a_j^2\delta_{\alpha\beta}
    \bigr].
\end{equation}

The Hamiltonian contribution is diagonal in this basis and takes the form
\begin{equation}
    (M_h)_{\alpha\beta}=-iE_\alpha\,\delta_{\alpha\beta}.
\end{equation}
Collecting all contributions, the full effective drift matrix is
\begin{equation}
    M = M_h + Q + \frac{1}{2}\sum_j A_j^2 .
    \label{eq:M_def_j}
\end{equation}
The term $\tfrac{1}{2}\sum_j A_j^2$ arises from It\^o calculus applied to multiplicative noise and represents the usual Stratonovich--It\^o correction~\cite{Pesce2013, reis2021relationstratonovichitointegrals} (see Appendix~\ref{app:ito_backaction}).

A direct evaluation shows that the leading $\mathcal{O}(\gamma)$ mode-dependent contributions from this term cancel exactly against the corresponding $\mathcal{O}(\gamma)$ contributions in the projected measurement drift $Q$ (see Appendix~\ref{app:diag_cancellation}). The only remaining large contribution is a mode-independent scalar shift generated by quadratic noise contractions, which does not affect localization properties.

Consequently, the residual mode-dependent drift is parametrically suppressed in the Zeno regime and scales as $\mathcal{O}(\varepsilon^2\gamma)$. This implies that the real parts of the eigenvalues of the effective drift matrix behave as
\begin{equation}
    \operatorname{Re}\lambda_{\min}(M)
    \sim \varepsilon^2\gamma,
    \label{eq:Zeno_gap}
\end{equation}
which defines the spectral gap governing the relaxation of fluctuations in the quantum Zeno regime.

Using the linear stochastic equation~\eqref{eq:lin_sde_c} together with standard It\^o estimates (see Appendix~\ref{app:fluct_estimate}), one obtains a uniform-in-time bound on the second moment of the fluctuation amplitudes:
\begin{equation}
    \sup_{t\ge0}\mathbb{E}\|c_\alpha(t)\|^2 \le C < \infty .
\end{equation}
This bound ensures that the fluctuation dynamics remains stochastically stable at all times.

Restricting attention to an approximately diagonal representation of the drift matrix $M$, each fluctuation mode $c_\alpha$ obeys an effective Ornstein--Uhlenbeck--type equation~\cite{Gardiner1985},
\begin{equation}
    d c_\alpha = -\lambda_\alpha c_\alpha\,dt + \sigma_\alpha\,dW,
\end{equation}
where the damping rate scales as $\lambda_\alpha \sim \varepsilon^2 \gamma$, and the effective noise strength satisfies $\sigma_\alpha^2 = \sum_j |u_{j,\alpha}|^2$, reflecting the reduction of multiple independent noise channels to a single effective Wiener process:
\begin{equation}
    \sigma_\alpha^2
    \sim \sum_j |u_{j,\alpha}|^2
    \sim \sum_j \gamma\,|\varphi_\alpha(j)|^2\,|\varphi_0(j)|^2 .
\end{equation}
Using the estimate $|\varphi_\alpha(j)|\sim \varepsilon$ within the localization core of $\varphi_0$, one finds $\sigma_\alpha^2\sim \varepsilon^2\gamma$. The stationary variance of each mode, therefore, satisfies $2\lambda_\alpha\,\mathbb{E}|c_\alpha|^2 \sim \sigma_\alpha^2$, implying
\begin{equation}
 \label{OU_balance}
    \mathbb{E}|c_\alpha|^2 = \mathcal{O}(1).
\end{equation}

Finally, passing back to real space, we use the core amplitude scaling $|\varphi_\alpha(j)|\sim \varepsilon$ together with the fact that only $\mathcal{O}(1)$ fluctuation modes contribute appreciably within the localization core, and obtain the pointwise bound for the fluctuation field [given by Eq.~\eqref{eq:fluct_field_expansion}],
\begin{equation}
    |\delta\psi_j|
    \lesssim
    \sum_{\alpha=1}^{\mathcal{O}(1)} |c_\alpha|\,|\varphi_\alpha(j)|
    =
    \mathcal{O}(\varepsilon).
    \label{eq:delta_psi_scaling}
\end{equation}
We have confirmed the validity of this bound numerically using QSD evolution (see Appendix~\ref{app:bound_check}).

This bound is the key output of the fluctuation field analysis and will subsequently be used in Sec.~\ref{eff_poten} to construct the effective potential and to place explicit bounds on its fluctuations.

\subsection{Effective potential}
\label{eff_poten}

We now use the separation of scales established above to construct the effective static potential associated with a typical QSD trajectory. This construction should be viewed as an instantaneous inverse problem: given an instantaneous state and the reference energy, we ask which site-dependent potential would make the state an eigenstate of a discrete Schr\"odinger equation. The resulting potential describes the spatial structure selected by the monitored dynamics. The purpose of this subsection is to show that, in the Zeno regime, this reconstructed potential reduces to the simple non-Hermitian form $V_j-i\gamma/2$, up to controlled corrections.

%We now show how the effective static potential and its fluctuations emerge from the QSD dynamics. 
Fixing a reference energy to $E_{\mathrm{dom}}$, we introduce an instantaneous QSD potential by construction.
Specifically, at each fixed time $t$ we define a site-dependent potential $V^{\mathrm{QSD}}_j(t)$ by requiring that the instantaneous state $\psi(t)$ satisfies a discrete Schr\"odinger eigenvalue equation:
\begin{equation}
    \label{eq:qsd_recurrence_repeat}
    -J\bigl(\psi_{j+1}(t)+\psi_{j-1}(t)\bigr)
    + V^{\mathrm{QSD}}_j(t)\,\psi_j(t)
    = E_{\mathrm{dom}}\,\psi_j(t).
\end{equation}
For sites where $\psi_j(t)\neq0$, this condition uniquely determines the potential, leading to the algebraically exact expression
\begin{equation}
    \label{eq:v_qsd_exact_repeat}
    V^{\mathrm{QSD}}_j(t)
    =
    \frac{E_{\mathrm{dom}}\psi_j(t)
	+J\bigl(\psi_{j+1}(t)+\psi_{j-1}(t)\bigr)}{\psi_j(t)} .
\end{equation}
In the Zeno regime, although the wavefunction evolves stochastically according to the QSD equation, it is nevertheless possible at each instant to reconstruct an effective potential $V_{\mathrm{QSD}}(t)$ for which the instantaneous state $\psi(t)$ is an eigenstate of a time-independent Schr\"odinger operator (with time treated as a parameter) with eigenvalue $E_{\mathrm{dom}}$.
This construction should not be interpreted as a statement about the actual dynamics.
Rather, it is an algebraic snapshot reconstruction: given $\psi(t)$, one simply asks which potential would render it an eigenstate with energy $E_{\mathrm{dom}}$.
Viewed in this way, the reconstructed potential $V_{\mathrm{QSD}}(t)$ provides a transparent and intuitive representation of how continuous measurement and backaction reshape the effective energy landscape experienced by a typical trajectory.

%The starting point is the exact algebraic identity obtained by inverting a static lattice Schr\"odinger equation at fixed time, Eq.~\eqref{eq:v_qsd_exact_repeat},
% \begin{equation}
% \label{eq:VQSD_exact_again}
% V^{\mathrm{QSD}}_j(t)
% =
% \frac{E_{\mathrm{dom}}\psi_j(t)
% + J\bigl(\psi_{j+1}(t)+\psi_{j-1}(t)\bigr)}{\psi_j(t)} ,
% \end{equation}
% \marcin{Seems to be a repeated Eq.~\ref{eq:v_qsd_exact_repeat}, we should probably just refer to it.}
%which defines an instantaneous effective potential associated with the wavefunction snapshot $\psi(t)$. We again point out that this identity is purely algebraic and does not describe the time evolution.

Eq.~\eqref{eq:v_qsd_exact_repeat} serves as the starting point for a controlled expansion around a localized reference state.
In the monitored AAH model, the dominant steady-state frequency vanishes, $E_{\mathrm{dom}}=0$, as per discussion in Sec.~\ref{subsec:dominant_energy}, so that the steady-state wavefunction is strictly time independent.
Using the decomposition from Eq.~\eqref{eq:inst_wfn_decomp} and since the fluctuation field [Eq.~\eqref{eq:delta_psi_scaling}] is $\mathcal{O}(\varepsilon), \varepsilon \ll 1$, the localized reference profile $\varphi_0(j)$ satisfies
\begin{equation}
    \psi_j(t) = \varphi_0(j),
    \qquad
    \dot{\psi}_j = 0 .
\end{equation}
Starting from the deterministic part of the QSD equation,
\begin{equation}
    \dot{\psi}_j
    =
    - i \sum_k (H_{AAH})_{jk}\psi_k
    - \frac{\gamma}{2}\psi_j
    + \gamma |\psi_j|^2 \psi_j ,
\end{equation}
with $(H_{AAH})_{jk}=-J(\delta_{j,k+1}+\delta_{j,k-1})+V_j\delta_{jk}$, and imposing $\dot{\psi}_j=0$, one obtains 
% the stationary balance condition
% \begin{equation}
%     - i \sum_k (H_{AAH})_{jk}\varphi_0(k)
%     - \frac{\gamma}{2}\varphi_0(j)
%     + \gamma |\varphi_0(j)|^2 \varphi_0(j)
%     = 0 .
% \end{equation}
% Writing the Hamiltonian terms explicitly and multiplying by $i$ yields
\begin{align}
    - J(\varphi_0&(j+1)+\varphi_0(j-1)) + V_j \varphi_0(j) \nonumber\\
    &= \frac{i\gamma}{2}\varphi_0(j)
    - i \gamma |\varphi_0(j)|^2\varphi_0(j)
    + \mathcal{O}(\varepsilon^2 J),
    \label{eq:phi0_equation}
\end{align}
which expresses the steady-state balance between coherent hopping, static disorder, and measurement-induced decay and nonlinear backaction. This equation is not a spectral eigenvalue problem but a self-consistency condition defining the dominant steady-state profile selected by the QSD dynamics.

Expanding the numerator and denominator of
Eq.~\eqref{eq:v_qsd_exact_repeat} to linear order in $\delta\psi$, and using
\begin{equation}
    (\varphi_0+\delta\psi)^{-1}
    =
    \varphi_0^{-1}\bigl(1-\delta\psi/\varphi_0\bigr)
    + \mathcal{O}(\delta\psi^2),    
\end{equation}
one obtains
\begin{align}
    V^{\mathrm{QSD}}_j(t) &\approx \frac{E_{\mathrm{dom}}\varphi_0(j) + J(\varphi_0(j+1)+\varphi_0(j-1))}{\varphi_0(j)} \nonumber\\
    &\quad + \frac{1}{\varphi_0(j)} \Bigl[ E_{\mathrm{dom}}\delta\psi_j + J(\delta\psi_{j+1}+\delta\psi_{j-1}) \Bigr] \nonumber\\
    &\quad - \frac{\delta\psi_j}{\varphi_0(j)^2} \Bigl[ E_{\mathrm{dom}}\varphi_0(j) {+} J(\varphi_0(j{+}1){+}\varphi_0(j{-}1)) \Bigr] \nonumber \\
    &\quad + \mathcal{O}(\delta\psi^2).
    \label{eq:V_expand}
\end{align}
Substituting the stationary balance condition~\eqref{eq:phi0_equation} into this
expression, the instantaneous potential can be written as
\begin{equation}
    \label{eq:VQSD_split}
    V^{\mathrm{QSD}}_j(t)
    =
    V_j
    - \frac{i\gamma}{2}
    + i \gamma |\varphi_0(j)|^2
    + \delta V_j(t),
\end{equation}
where $\delta V_j(t)$ collects all fluctuation-dependent contributions. We emphasize that the effective non-Hermitian operator associated with Eq.~\eqref{eq:VQSD_split} does not generate the time evolution of the QSD dynamics; rather, it is a static construct obtained from instantaneous wavefunction snapshots and is used solely to characterize the spatial structure of typical QSD trajectories.

Expanding $\delta\psi_j$ in the orthogonal fluctuation modes [Eq.~\eqref{eq:fluct_field_expansion}], and retaining only
terms linear in the amplitudes $c_\alpha$, one finds
\begin{equation}
    \label{eq:deltaV_def}
    \delta V_j(t)
    =
    \frac{1}{\varphi_0(j)}
    \sum_{\alpha\ge1}
    c_\alpha(t)\,(E_\alpha-E_{\mathrm{dom}})\,\varphi_\alpha(j)
    + \mathcal{O}(\delta\psi^2).
\end{equation}
The term $i\gamma|\varphi_0(j)|^2$ in Eq.~\eqref{eq:VQSD_split} is nonzero only on $\mathcal{O}(1)$ lattice sites due to the strong localization of $\varphi_0$, and therefore does not contribute to the Lyapunov exponent in the thermodynamic limit. Using $|E_\alpha-E_{\mathrm{dom}}|=\mathcal{O}(J)$ together with $|\varphi_\alpha(j)|=\mathcal{O}(\varepsilon)$, one concludes that
\begin{equation}
    V^{\mathrm{QSD}}_j(t)
    =
    V_j
    - \frac{i\gamma}{2}
    + \mathcal{O}\!\left(
    \frac{J^2}{\sqrt{\lambda^2+(\gamma/2)^2}}
    \right)\!,
    \label{eq:eff_pot_qsd}
\end{equation}
where the deterministic correction (after diagonal cancellation) satisfies $\Delta V_j^{\mathrm{det}}=\mathcal{O}(J\varepsilon)$. The remaining stochastic fluctuations are parametrically small and do not affect the leading-order localization length.

The effective potential of $V_j - i \gamma/2$ from Eq.~\eqref{eq:eff_pot_qsd} is the central result of the effective theory construction: the monitored trajectory behaves, for localization purposes, as if it were governed by the original quasiperiodic potential supplemented by a uniform imaginary measurement-induced term, plus controlled fluctuation corrections.
This effective potential (without corrections), together with the corresponding non-Hermitian Hamiltonian, has been discussed in the literature primarily in the context of postselected dynamics~\cite{Turkeshi2021, 10.21468/SciPostPhys.14.3.031, 10.21468/SciPostPhys.14.5.138, DeTomasi2024}, for instance when considering quantum jumps instead of QSD and postselecting upon the no-click limit. However, it is not a priori evident that postselected trajectories faithfully capture the properties of generic quantum trajectories. In contrast, our analysis does not rely on postselection and explicitly gives the order of the correction terms in the Zeno limit for the full QSD evolution.

\subsection{Localization length}

With the effective potential in hand, we finally translate the QSD problem into the language of the localization theory. The transfer-matrix formulation converts the spatial decay of a wavefunction into a Lyapunov exponent, whose inverse gives the localization length. The purpose of this subsection is to show that the transfer matrices obtained from full QSD trajectories differ from those of the effective static non-Hermitian model only by perturbations controlled by the parameter $\varepsilon$. Consequently, the corresponding Lyapunov exponents, and hence localization lengths, agree up to $\mathcal{O}(\varepsilon^2)$ corrections.

%In the following, the Lyapunov exponent is evaluated at the dominant energy $E_{\mathrm{dom}}$, with stochastic energy fluctuations contributing only subleading, self-averaging corrections. We analyze how stochastic fluctuations generated by quantum state diffusion modify the localization properties encoded in the transfer-matrix formulation.
Fixing the dominant energy $E_{\mathrm{dom}}$, the discrete single-particle Schr\"odinger equation may be written in transfer-matrix form as
\begin{equation}
    \begin{pmatrix}
    \psi_{j+1}\\
    \psi_j
    \end{pmatrix}
    =
    T_j(E_{\mathrm{dom}})
    \begin{pmatrix}
    \psi_j\\
    \psi_{j-1}
    \end{pmatrix},
\end{equation}
where the effective transfer matrix associated with the leading, time-independent QSD potential $V^{\mathrm{eff}}_j$ is
\begin{equation}
    T^{\mathrm{eff}}_j(E_{\mathrm{dom}})
    =
    \begin{pmatrix}
    \dfrac{E_{\mathrm{dom}}-V^{\mathrm{eff}}_j}{J} & -1\\[6pt]
    1 & 0
    \end{pmatrix}.
\end{equation}
Along an individual QSD trajectory, the instantaneous potential fluctuates around its effective value according to
\begin{equation}
    V^{\mathrm{QSD}}_j(t)=V^{\mathrm{eff}}_j+\delta V_j(t),
\end{equation}
where the fluctuation $\delta V_j(t)$ is induced by the small wavefunction correction $\delta\psi_j(t)$. Using the pointwise bound from Eq.~\eqref{eq:delta_psi_scaling} and linearizing the ratios of neighboring amplitudes entering the transfer matrix, the instantaneous QSD transfer matrix may be written as
\begin{equation}
    T^{\mathrm{QSD}}_j(t)=T^{\mathrm{eff}}_j+\delta T_j(t),
\end{equation}
where $\delta T_j(t)$ is linear in $\delta V_j(t) / J$ and therefore of order $\varepsilon$. %\marcin{I've added ``$/J$'', please check if that's fine. Otherwise, it could imply that $\delta V_j(t)$ was of order $\varepsilon$.} 
More precisely, for any submultiplicative matrix norm,
\begin{equation}
    \mathbb{E}\,\|\delta T_j\|^2
    =
    \mathcal{O}(\varepsilon^2)
    =
    \mathcal{O}\!\left(\frac{J^2}{\lambda^2+(\gamma/2)^2}\right).
\end{equation}
The stochastic process $\{\delta T_j(t)\}$ is stationary and possesses finite second moments, ensuring that products of the corresponding random transfer matrices satisfy the standard assumptions required for the existence and stability of Lyapunov exponents.

The localization properties are governed by the largest Lyapunov exponent:
\begin{equation}
    \kappa
    =
    \lim_{n\to\infty}
    \frac{1}{n}
    \log
    \Big\|
    \prod_{j=1}^{n} T_j
    \Big\|,
    \label{eq:kappa_definition}
\end{equation}
which is defined almost surely and independently of the chosen matrix norm.
Denoting by $\kappa_{\mathrm{eff}}$ the Lyapunov exponent associated with the deterministic product of matrices $T^{\mathrm{eff}}_j$, and by $\kappa_{\mathrm{QSD}}$ the corresponding exponent for the random matrices $T^{\mathrm{QSD}}_j(t)$, continuity of the largest Lyapunov exponent under perturbations of the matrix distribution implies
\begin{equation}
    \kappa_{\mathrm{QSD}}
    =
    \kappa_{\mathrm{eff}}
    +
    \mathcal{O}(\varepsilon^2).
    \label{eq:kappa_eff}
\end{equation}
The absence of a linear correction reflects the fact that the QSD-induced fluctuations are centered, so that the leading contribution arises at second order in the perturbation strength.

In one dimension, the Lyapunov exponent is the inverse localization length, $\xi=\kappa^{-1}$. Expanding in $\varepsilon$ therefore yields
\begin{equation}
    \xi_{\mathrm{QSD}}
    =
    \xi_{\mathrm{eff}}
    +
    \mathcal{O}(\varepsilon^2),
    \label{eq:xi_eff}
\end{equation}
demonstrating that the localization length extracted from the full QSD dynamics coincides with that of the effective non-Hermitian description up to parametrically small corrections controlled by $J^2/[\lambda^2+(\gamma/2)^2]$ in the quantum Zeno regime.

Eqs.~\eqref{eq:kappa_eff} and \eqref{eq:xi_eff} are the main analytical conclusions of Sec.~\ref{sec:qsd_heff_equivalence}: they justify comparing the localization length and the Lyapunov exponent extracted from full QSD trajectories directly with those obtained from the effective non-Hermitian transfer-matrix problem.

Overall, the derivation for the localization length in Sec.~\ref{sec:qsd_heff_equivalence} is more broadly applicable and not restricted to the monitored AAH model. Any free-fermion Hamiltonian with the same nearest-neighbor hopping term, but a different potential term $V_j$ should admit a similar description in the Zeno regime: the main quantities that would change are the dominant energy $E_{\mathrm{dom}}$, and the parameter $\varepsilon$ (which instead becomes a different function of the microscopic model parameters). The effective potential would remain $V_j^{\mathrm{eff}} = V_j - i\gamma/2$, and the resulting localization length from the transfer-matrix approach would again receive only $\mathcal{O}(\varepsilon^2)$ corrections. In this sense, the correspondence between QSD dynamics and this effective Zeno theory provides a general route to extracting localization properties in monitored free-fermion systems.

%\section{Results}
\section{Localization length in the quantum Zeno regime}
\label{sec:results}

%\subsection{Analytical regimes and comparison with numerics}
%\label{subsec:results_analytics}
	
% 	We now summarize the analytical predictions for the Lyapunov exponent
% $\kappa$ in relevant parameter limits and compare them with localization
% lengths extracted numerically from steady-state QSD trajectories (after orbital
% unscrambling). The analytical expressions follow from the effective static
% description in the Zeno limit and from known results for the bare
% Aubry--Andr\'e--Harper (AAH) model; the localization length is obtained from the
% corresponding transfer-matrix Lyapunov exponent. Numerical localization lengths are
% computed by averaging logarithmic decay profiles over left and right tails, orbitals,
% and independent QSD realizations (see Sec.~\ref{subsec:loc_length}).
\begin{figure}[t]
    \centering
    \includegraphics[width=0.9\linewidth]{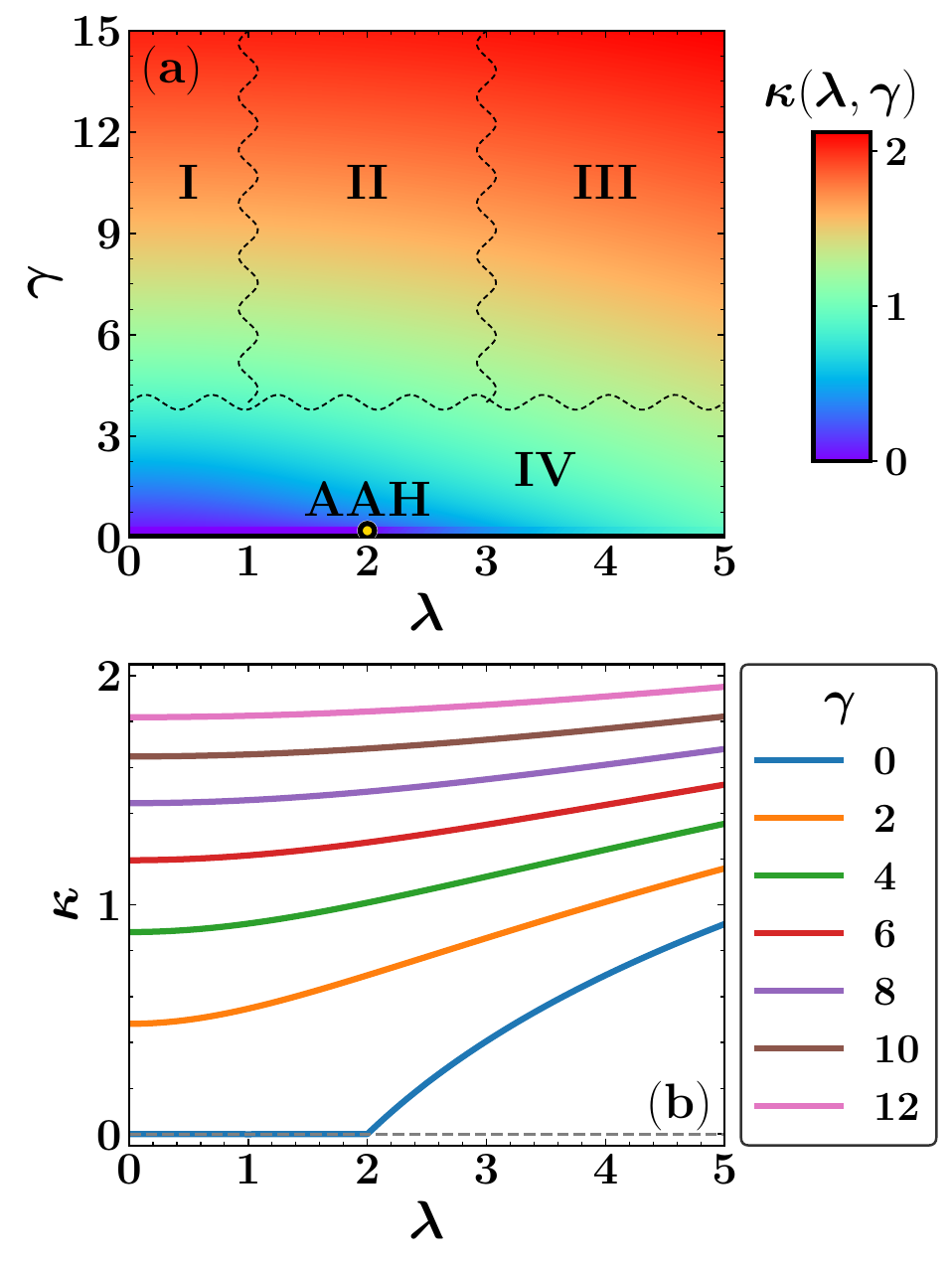}
    \caption{
    Effective-theory Lyapunov exponent $\kappa$ in the $\lambda$--$\gamma$ parameter space for $J=1$.
    (a)~Density plot of $\kappa(\lambda,\gamma)$ showing approximate regimes separated by wiggly black dashed lines indicating gradual crossovers rather than sharp transitions.
    \textbf{Regimes I, II, III} ($\gamma \gtrsim 4$): Zeno (measurement-dominated) regime with varying quasiperiodic contributions.
    \textbf{Regime I} ($\lambda \lesssim 1$): weak quasiperiodic effect, dynamics primarily governed by measurements.
    \textbf{Regime II} ($1 \lesssim \lambda \lesssim 3$): intermediate crossover where measurement and quasiperiodic effects compete.
    \textbf{Regime III} ($\lambda \gtrsim 3$): strong quasiperiodic coupling dominates even with strong measurements.
    \textbf{Regime IV} ($\gamma \lesssim 4$): weak measurement regime with dominant quasiperiodic localization.
    %The wavy boundaries emphasize that transitions between regimes are continuous and lack strict phase boundaries.
    The wavy dashed boundaries emphasize that the changes between regimes occur smoothly and do not correspond to sharply defined separations.
    The gold circle at $\lambda=2$, $\gamma = 0$ marks the AAH critical point, and the line at $\gamma=0$ corresponds to the unmonitored AAH model.
    (b)~$\kappa$ versus $\lambda$ for fixed $\gamma$ values.
    }
    \label{fig:phase_kappa_lambda}
\end{figure}

\begin{figure*}[t]
	\centering
	\includegraphics[width=\textwidth]{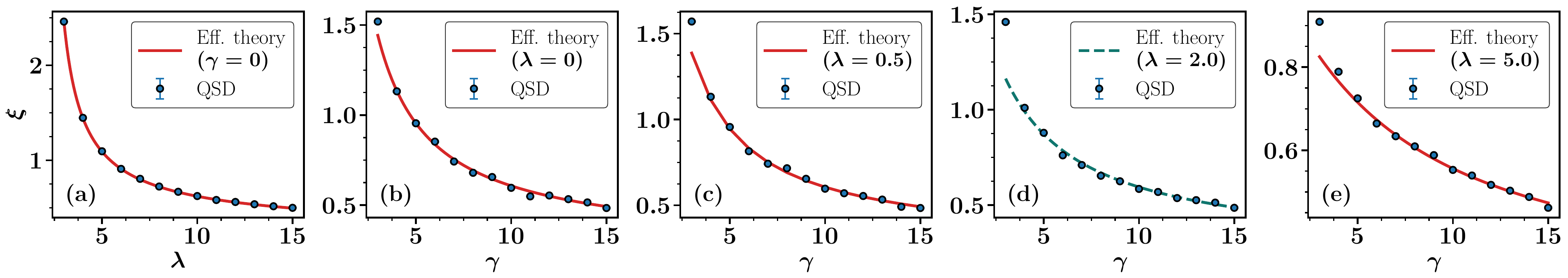}
	\caption{
		Localization length $\xi$ as a function of the measurement rate $\gamma$ and strength of the quasiperiodic potential $\lambda$, with $J=1$ in all cases.
		(a)~$\gamma = 0$, the unmonitored AAH model.
		(b)~$\lambda = 0$, corresponding to purely measurement-induced (Zeno) localization.
		(c)~$\lambda = 0.5$, representing the measurement-dominated regime with a weak quasiperiodic potential, where localization is primarily governed by measurement backaction.
        (d)~$\lambda = 2.0$, a crossover regime with intermediate potential strength.
		(e)~$\lambda = 5.0$, corresponding to the strong-potential regime, where intrinsic localization due to the quasiperiodic potential dominates the dynamics.
		In all panels, symbols denote numerical results obtained from left--right averaged orbitals after QSD evolution, while lines indicate the corresponding theoretical predictions from the effective theory. In panel~(d), the effective theory is calculated numerically using the transfer-matrix approach (green dashed line), whereas the other panels use closed-form expressions found in the text (red solid lines).
        Plotted error bars (smaller than the markers) indicate the standard error across trajectories.
    }
	\label{fig:xi_four_panels}
\end{figure*}

We now confront the effective static description in the Zeno regime developed in the previous section with numerical results obtained from the full QSD dynamics. The Lyapunov exponents are computed using the transfer-matrix method with a total iteration count of $20\,000$ (the initial transient behavior is discarded to ensure convergence and numerical stability). The QSD numerics is obtained from the steady-state QSD trajectories after orbital unscrambling and is averaged over 96 independent noise realizations, as described in Sec.~\ref{subsec:loc_length}, for a system size of $L=256$. The time evolution is performed up to $t_{\text{final}}=1200$ with time step $dt=0.01$, while the unscrambling procedure precision is $10^{-24}$. 

Fig.~\ref{fig:phase_kappa_lambda} shows the Lyapunov exponent $\kappa$, computed from the effective theory using the transfer-matrix approach, across a range of quasiperiodic potential strengths $\lambda$ and measurement strengths $\gamma$. This parameter space naturally separates into several regimes, each characterized by a distinct balance between coherent hopping, quasiperiodicity, and measurement backaction; in several of these regimes, closed-form expressions for $\kappa$ can be derived (see Appendix~\ref{app:transfer_matrix} for the complete derivations). These regimes do not represent distinct physical phases; rather, they characterize limiting behaviors of the system. Regimes I--III correspond to the Zeno regime, defined operationally by the excellent agreement between the effective theory and QSD results (specifically, $|\xi_{\mathrm{QSD}}-\xi_{\mathrm{eff}}| \lesssim 0.025$, which holds approximately for $\gamma \gtrsim 4$). The remaining region, $\gamma \lesssim 4$, constitutes the weak-measurement Regime IV. Within the Zeno regime, we further distinguish three representative cases: Regime I, where measurement dominates over the quasiperiodic potential (the Lyapunov exponent is captured well by a closed-form expression to within $\approx 0.1\%$, or about $\lambda \lesssim 1$); Regime III, where both quasiperiodicity and measurement strength dominate (the corresponding closed-form expression is accurate to within $\approx 4\%$, or $\lambda \gtrsim 3$); and Regime II, which covers the intermediate-disorder window $1 \lesssim \lambda \lesssim 3$. Below, we discuss each regime, together with the accompanying QSD and effective theory results.

%\vspace{0.5em}
%\noindent
\subsection{Aubry--Andr\'e--Harper reference case (\texorpdfstring{\(\gamma=0\)}{})}
In the absence of measurements, the system reduces to the standard AAH model. The Lyapunov exponent is given by the well-known exact result
\begin{equation}
    \kappa_{\mathrm{AAH}} =
    \begin{cases}
    0, & |\lambda| \le 2J,\\[4pt]
    \ln\!\left(\dfrac{|\lambda|}{2J}\right), & |\lambda|>2J,
    \end{cases}
\end{equation}
implying extended eigenstates for $|\lambda|\le 2J$ and exponential localization for $|\lambda|>2J$, with critical points at $|\lambda|=2J$. Fig.~\ref{fig:xi_four_panels}(a) shows the numerically extracted Lyapunov exponents (symbols) together with the analytic prediction (solid line). The agreement in the localized phase is excellent, and we expect it to persist until $\xi$ becomes comparable with the system size ($|\lambda| \sim 2Je^{1/L}$), i.e., when finite-size effects near the critical point become important.%, while deviations near $|\lambda|=2J$ reflect finite-size effects and the expected crossover behavior close to the critical point. %\marcin{We don't see any finite-size effects on the plot. Maybe we can omit talking about FSE for the clean AAH model, since it has a known critical behaviour?}
%\pinaki{At $\lambda = 3$, also does not give any finite size effect..
%}
%\marcin{In that case, how about adding point $\lambda=3$ in Fig. 4(a)?} \pinaki{ok i will add this}

\subsection{Regime I -- Measurement-dominated (quantum Zeno) regime (\texorpdfstring{\(\gamma \gg J\)}{} and \texorpdfstring{\(0\le\lambda\ll\gamma\)}{})}

We first consider the regime in which continuous measurement dominates the dynamics, \(\gamma \gg J\). In the absence of any static potential (\(\lambda=0\)), the effective static description yields the Lyapunov exponent
\begin{equation}
    \kappa(0,\gamma)
    \approx
    \operatorname{arsinh}\!\left(\frac{\gamma}{4J}\right),
\end{equation}
which asymptotically approaches $\ln[\gamma/(2J)]$ for \(\gamma \gg J\). Numerically extracted Lyapunov exponents from the QSD trajectories follow this prediction closely throughout the Zeno regime, as shown in Fig.~\ref{fig:xi_four_panels}(b). Small deviations at intermediate values of \(\gamma \lesssim 4\) arise from finite-\(J/\gamma\) corrections and finite-size effects.

When a weak quasiperiodic potential is present (\(\lambda \ll \gamma\)), measurements continue to set the dominant localization mechanism. In this measurement-dominated limit, the Lyapunov exponent admits a perturbative expansion about the Zeno result,
% \begin{equation}
% \kappa(\lambda,\gamma)
% =
% \operatorname{arsinh}\!\left(\frac{\gamma}{4J}\right)
% +
% \mathcal{O}\!\left(\frac{\lambda^2}{\gamma^2}\right).
% \end{equation}
\begin{equation}
    \kappa(\lambda,\gamma)
    \approx
    \operatorname{arsinh}\!\left(\frac{\gamma}{4J}\right)
    +
    \frac{\lambda^2}{\gamma^2}.
\end{equation}
Thus, the leading localization length term is entirely controlled by the measurement rate, with the quasiperiodic potential contributing only a subleading correction. Our numerical data [see Fig.~\ref{fig:xi_four_panels}(c)] confirm that the dominant dependence of \(\kappa\) is set by \(\gamma\), and that the residual discrepancy between numerical and analytic results decreases systematically with increasing \(\gamma\), scaling approximately as $\mathcal{O}(\varepsilon^2) \approx (J / \gamma)^2$ in the large-\(\gamma\) regime.% and the error analysis in Fig.~\ref{fig:compare}(d)).

\subsection{Regime II -- Measurement-dominated intermediate regime (\texorpdfstring{$\lambda \sim J$}{}, \texorpdfstring{$\gamma \gg J$}{})}

We next consider an intermediate regime in which the quasiperiodic potential is of the order of the hopping amplitude, while the dynamics is strongly dominated by measurement, \(\gamma \gg J\). In this regime, no closed-form expression for the Lyapunov exponent is available. Instead, \(\kappa\) must be obtained numerically from the transfer-matrix formulation of the effective static model. We then compare these results to the steady-state QSD numerics in Fig.~\ref{fig:xi_four_panels}(d) for an intermediate value of $\lambda$. The two approaches yield consistent localization lengths, with \(\kappa\) interpolating smoothly between the pure Zeno limit and the strong-potential regime as \(\lambda\) is increased. Measurement remains the dominant localization mechanism, while the quasiperiodic potential provides a subleading enhancement.%\pinaki{I changed this region in  the fig.3 plot please do check is this valid or not ??} \marcin{How about we don't explicitly write the region of $\lambda$, but say $\lambda \sim J$? The boundaries for these regions aren't precisely defined, rather these are some loosely-defined regimes.}
\begin{table*}[t]
	\centering
	\caption{Asymptotic expressions for the Lyapunov exponent $\kappa(\lambda,\gamma)$ at the dominant energy scale $E_{\mathrm{dom}}=0$ in the monitored Aubry--Andr\'e--Harper chain, extracted from the effective theory.}
	\label{tab:lyap_allregions}
	\setlength{\tabcolsep}{12pt}
	\renewcommand{\arraystretch}{2.1}
	\begin{tabular}{l l l}
		\hline\hline
		Physical regime
		& Parameter conditions
		& Asymptotic Lyapunov exponent $\kappa(\lambda,\gamma)$ \\ 
		\hline
		
		Unmonitored AAH
		& $\gamma=0$
		&
		$\displaystyle
		\kappa_{\mathrm{AAH}}(\lambda)=
		\begin{cases}
		0, & |\lambda|\le 2J,\\
		\ln\!\left|\dfrac{\lambda}{2J}\right|, & |\lambda|>2J
		\end{cases}
		$ \\[8pt]
		
		Measurement-dominated (Zeno)
		& $\gamma\gg J,\ \lambda\ll\gamma$
		&
		$\displaystyle
		\kappa(\lambda,\gamma)
		\approx
		\operatorname{arsinh}\!\left(\frac{\gamma}{4J}\right)
		+
		\frac{\lambda^2}{\gamma^2}
        + \mathcal{O}\!\left(\frac{J^2\lambda^2}{\gamma^4},\frac{\lambda^4}{\gamma^4}\right)
		$ \\[8pt]
		
		Strong coupling
		& $\min(\lambda,\gamma)\gg J$
		&
		$\displaystyle
		\kappa(\lambda,\gamma)
		\approx
		\ln\!\left|\frac{\lambda}{2J}\right|
		+
		\operatorname{arsinh}\!\left(\frac{\gamma}{2\lambda}\right)
        + \mathcal{O}\!\left(\frac{J\varepsilon}{\gamma}\right)
		$ \\
		
		\hline\hline
	\end{tabular}
\end{table*}

\subsection{Regime III -- Cooperative strong-coupling localization (\texorpdfstring{\(\lambda,\gamma\gg J\)}{})}

When both the potential and measurement strength are large, the two localization mechanisms act cooperatively. In this limit, the Lyapunov exponent is well approximated by the additive form
\begin{equation}
    \kappa(\lambda,\gamma) \approx
    \ln\!\left(\frac{|\lambda|}{2J}\right) + \operatorname{arsinh}\!\left(\frac{\gamma}{2\lambda}\right),
\end{equation}
which combines the static AAH contribution with a measurement-induced term evaluated relative to the potential scale. As shown in Fig.~\ref{fig:xi_four_panels}(e), the numerically extracted \(\kappa\) follows this additive trend; small systematic deviations can be attributed to subleading $\mathcal{O}(\varepsilon^2)$ corrections and finite-size crossover effects.

\subsection{Regime IV -- Weak measurement regime (\texorpdfstring{$\gamma \lesssim J$}{})}

Outside the Zeno regime, the effective theory is no longer expected to hold, as the subleading corrections proportional to $\varepsilon^2 = J^2/[\lambda^2+(\gamma/2)^2]$ become non-negligible. In Fig.~\ref{fig:xi_four_panels}, we indeed observe that the predictions of the effective theory begin to deviate from the QSD results once $\gamma \lesssim 4$. It remains unclear whether the QSD Lyapunov exponent behaves continuously as one approaches the unmonitored limit. Studies of other disordered free-fermion models~\cite{PhysRevB.108.165126, PhysRevB.110.024303} indicate that a discontinuity may arise between the unmonitored case $\gamma = 0$ (corresponding here to the pure AAH model) and the weak-measurement regime $\gamma > 0$. Moreover, although the numerical results in this limit may suggest a critical phase~\cite{3zfd-3hqt}, the generic result of Ref.~\cite{PhysRevX.13.041046} shows that localization will always occur for any $\gamma > 0$ in this system, with correlation length $\propto (J/\gamma) \exp(\sqrt{2} \pi J / \gamma)$ in the limit of $\gamma \to 0$ and for small $\lambda$.

\subsection{Overall comparison and remarks}
Across the parameter space, the effective theory captures the dominant scaling of \(\kappa\) and correctly predicts the asymptotic Zeno behavior. Table~\ref{tab:lyap_allregions} presents the summary of the asymptotic expressions for the Lyapunov exponent extracted from the effective theory. Quantitatively, the QSD-based localization lengths agree with the effective static description to within the expected perturbative corrections; in particular, discrepancies in the Zeno regime scale parametrically as \(\mathcal{O}(J^2/[\lambda^2+(\gamma/2)^2])\). 
The strong measurement-induced localization (the quantum Zeno effect) couples with the intrinsic quasiperiodic localization to confine the fermions even more effectively, thereby strongly suppressing quantum information transport.
% Likewise, the spreading of entanglement should be simultaneously hindered by the quasiperiodic potential and by continuous monitoring. UNSURE
The overall agreement observed in the Zeno regime demonstrates that the effective static framework offers a robust and physically insightful approach for characterizing localization in monitored disordered systems.
% \marcin{I think the following clean AAH discussion isn't as interesting}
%Finite-size crossovers are most pronounced near the AAH critical point \(|\lambda|=2J\), where the apparent critical behavior drifts with system size (see inset of Fig.~\ref{fig:compare}(a)). Figures~\ref{fig:compare}(a)--(e) summarize the detailed numerical versus analytical comparisons discussed above.

\section{Conclusion}
\label{sec:conclusion}

We have investigated the localization properties of a continuously monitored free-fermionic system subject to a quasiperiodic potential, a setting that allows for substantial analytic treatment. In the Zeno regime, strong measurements suppress coherent transport and stabilize localized spatial profiles of single-particle wave functions. We devise an effective static description of the monitored dynamics (that does not rely on postselection) and find that, despite the stochastic nature of individual trajectories, the dominant localization features are governed by an effective static potential whose leading contribution coincides with that of a simple non-Hermitian Hamiltonian. Our effective theory predictions for the localization length match the numerical results from quantum state diffusion across the Zeno regime, with only small, systematically diminishing corrections as the measurement strength increases. This, in turn, enables direct control over the localization length (and consequently over the decay of two-point correlation functions and quantum mutual information) by tuning the measurement strength and quasiperiodicity. With these results, we establish a direct, quantitative connection between full stochastic monitored dynamics and non-Hermitian localization theory, validating the latter as a practical and accurate description of measurement-induced localization.

Our results open several promising directions for future work.
A natural next step is to extend the present analysis beyond noninteracting fermions. While measurement-induced transitions in many-body localized systems have been explored previously~\cite{Lunt2020, Boorman2022, Tang2025}, the specifics of the localization theory in the Zeno regime remain poorly understood; introducing quasiperiodicity into the disordered potential may offer a route toward analytical progress.
Another important question concerns the role of global symmetries: the effective field theory of the transition in monitored free fermions is known to depend sensitively on conservation laws~\cite{PhysRevX.13.041046, Turkeshi2021}, and it would be valuable to clarify how these constraints reshape the emergent Zeno-limit dynamics.
It is also worth examining how non-Hermitian spectral features, such as exceptional points and mobility edges, manifest in effective descriptions of monitored systems and how they relate to measurement-induced criticality.
In general, establishing deeper connections between stochastic quantum-trajectory evolution and effective models in the Zeno regime would not only sharpen our conceptual understanding of monitored dynamics but also help identify new opportunities for harnessing measurement as a tool for quantum control and state engineering.

% \section{Acknowledgements}

\begin{acknowledgments}
We thank Sumilan Banerjee and Vatsana Tiwari for valuable discussions and support. 
P.~S. acknowledges IISER Bhopal for the Ph.D. fellowship. N.~R. acknowledges partial support through NISER Plan Project No. RIN 4001-SPS.
M.~S. was supported by the Engineering and Physical Sciences Research Council (EPSRC) grant on Robust and Reliable Quantum Computing (RoaRQ), Investigation 004 (grant reference EP/W032635/1), and the EPSRC grant on (De)constructing quantum software (DeQS) (grant reference EP/Z002230/1).
We are grateful to the High Performance Computing (HPC) facility at IISER Bhopal, where the large-scale calculations in this project were performed.
The authors would also like to acknowledge the use of the University of Oxford Advanced Research Computing (ARC) facility in carrying out this work. \url{http://dx.doi.org/10.5281/zenodo.22558}.

The data that support the findings of this article are openly available at~\cite{Data}.
\end{acknowledgments}

% \section{Appendix}

\appendix

\section{Scaling of energy fluctuations with system size}
\label{app:self_averaging}

In this appendix, we provide a general argument for the scaling of instantaneous energy fluctuations with system size. The key observation is that the instantaneous energy per particle is a spatial average over many weakly correlated local energy densities.

\paragraph{Instantaneous energy.}
Along a single quantum state diffusion (QSD) trajectory, the instantaneous energy per particle is defined as
\begin{equation}
    E_{\mathrm{inst}}(t)
    =
    \frac{1}{N}
    \langle \psi(t) | H | \psi(t) \rangle,
\end{equation}
where $N=L/2$ is the number of particles.
Since the Hamiltonian can be written as a sum of local terms,
\begin{equation}
    H = \sum_{i=1}^{L} h_i,
\end{equation}
we may express
\begin{equation}
    E_{\mathrm{inst}}(t)
    =
    \frac{1}{N}
    \sum_{i=1}^{L}
    \epsilon_i(t),
\end{equation}
where
\begin{equation}
    \epsilon_i(t)
    =
    \langle \psi(t) | h_i | \psi(t) \rangle
\end{equation}
denotes the local energy density at lattice site $i$. Each $\epsilon_i(t)$ is an $\mathcal{O}(1)$ stochastic quantity that fluctuates in time due to measurement backaction.

\paragraph{Variance of the instantaneous energy.}
We consider the variance over stochastic realizations (or, equivalently, in the stationary regime, over time along a single trajectory):
\begin{equation}
    \mathrm{Var}(E_{\mathrm{inst}})
    =
    \mathrm{Var}\!\left(
    \frac{1}{N}
    \sum_{i=1}^{L}
    \epsilon_i
    \right)
    =
    \frac{1}{N^2}
    \sum_{i,j=1}^{L}
    \mathrm{Cov}(\epsilon_i,\epsilon_j),
\end{equation}
where
\begin{equation}
    \mathrm{Cov}(\epsilon_i,\epsilon_j)
    =
    \langle \epsilon_i \epsilon_j \rangle_{\mathrm{traj}}
    -
    \langle \epsilon_i \rangle_{\mathrm{traj}}
    \langle \epsilon_j \rangle_{\mathrm{traj}}
\end{equation}
denotes the covariance between local energy densities, and averages $\langle \cdot \rangle_{\mathrm{traj}}$ are taken over quantum trajectories.

\paragraph{Self-averaging and scaling.}
In the Zeno-localized regime, correlations of local observables are
short-ranged: the connected covariance
$\mathrm{Cov}(\epsilon_i,\epsilon_j)$ decays rapidly with the
spatial separation $|i-j|$. Equivalently, the covariance function is
absolutely summable,
\begin{equation}
    \sup_i
    \sum_{j=1}^{L}
    \big|
    \mathrm{Cov}(\epsilon_i,\epsilon_j)
    \big|
    < \infty.
\end{equation}
As a result, for each fixed $i$, the inner sum
$\sum_j \mathrm{Cov}(\epsilon_i,\epsilon_j)$ is of order unity.
Summing over all $i$ then yields
\begin{equation}
    \sum_{i,j=1}^{L}
    \mathrm{Cov}(\epsilon_i,\epsilon_j)
    \sim \mathcal{O}(L).
\end{equation}
Since $N=L/2$, this scaling is equivalently $\mathcal{O}(N)$.

Substituting into the expression above gives
\begin{equation}
    \mathrm{Var}(E_{\mathrm{inst}})
    \sim
    \frac{L}{N^2}
    \sim
    \frac{1}{N},
\end{equation}
so that the standard deviation scales as
\begin{equation}
 \label{Fluc}
    \sigma_E
    =
    \sqrt{\mathrm{Var}(E_{\mathrm{inst}})}
    \sim
    N^{-1/2}.
\end{equation}
This reflects the self-averaging character of the instantaneous energy:
fluctuations are suppressed by spatial averaging over many weakly
correlated local contributions.

\paragraph{Numerical evaluation.}
In practice, the variance is obtained from the stationary time series
$\{E_{\mathrm{inst}}(t_k)\}$ via the sample variance
\begin{equation}
    \mathrm{Var}(E_{\mathrm{inst}})
    =
    \frac{1}{M}
    \sum_{k=1}^{M}
    \left[
    E_{\mathrm{inst}}(t_k)
    -
    \overline{E}_{\mathrm{inst}}
    \right]^2,
\end{equation}
which coincides with the ensemble variance under ergodicity of the
stationary QSD dynamics.

\paragraph{Clean limit.}
In the clean case ($\lambda=0$), the Hamiltonian and the stochastic
evolution are translationally invariant in distribution. Consequently,
ensemble-averaged local energy densities are identical across sites.
The suppression of fluctuations in this limit is therefore governed by
symmetry and the absence of spatial inhomogeneity, rather than by
self-averaging over disorder-induced variations.

% ============================================================
\section{Localization properties in the Zeno limit}
\label{app:localization_proof}

In this appendix, we derive the localization properties of the dominant mode $\varphi_0$ and of the fluctuation modes $\{\varphi_\alpha\}_{\alpha\ge1}$ in the Zeno regime. All statements are understood pathwise, i.e., for a fixed realization of the quantum state diffusion (QSD) trajectory at a given time. Once conditioned on a single trajectory, $\psi(t)$ is an ordinary Hilbert-space vector, and the problem reduces to deterministic operator equations, analogous to localization in systems with quenched disorder.

Throughout, we assume the Zeno hierarchy
\begin{equation}
    \min(\lambda,\gamma) \gg J,
    \qquad
    \varepsilon \equiv \frac{J}{\sqrt{\lambda^2+(\gamma/2)^2}} \ll 1,
    \label{eq:zeno_condition}
\end{equation}
which ensures strong diagonal dominance, i.e., diagonal terms dominate in the stochastic Schr\"odinger equation.% \marcin{How about this?}\pinaki{Its fine, but should I maintain it, where it more understandable?}\marcin{I think since we define it here, then it's fine to leave the rest as is.}\pinaki{Alright, let’s review the appendix first; afterwards, we can improve its readability.}

% ============================================================
\subsection*{A. Long-time stationary solution of the deterministic QSD flow}
% ============================================================

The deterministic part of the QSD equation in the site basis is
\begin{equation}
\dot{\psi}_j
=
-i\Big[-J(\psi_{j+1}+\psi_{j-1}) + V_j\psi_j\Big]
-\frac{\gamma}{2}\psi_j
+\gamma|\psi_j|^2\psi_j .
\label{eq:qsd_site}
\end{equation}

In the Zeno regime, the dominant component becomes stationary up to a phase. We therefore write
\begin{equation}
\psi_j(t) = e^{-iE_{\mathrm{dom}} t}\,\varphi_0(j),
\end{equation}
so that $\dot{\psi}_j = -iE_{\mathrm{dom}}\psi_j$. Substituting into Eq.~\eqref{eq:qsd_site} and dividing by $-i$ gives
\begin{align}
E_{\mathrm{dom}}\varphi_0(j)
&=
-J\bigl(\varphi_0(j+1)+\varphi_0(j-1)\bigr)
\nonumber\\
&\quad
+ \Bigl(
V_j
- \frac{i\gamma}{2}
+ \gamma|\varphi_0(j)|^2
\Bigr)\varphi_0(j).
\label{eq:dominant_equation}
\end{align}

Defining the discrete Laplacian
\begin{equation}
(\Delta \varphi)_j \equiv \varphi(j+1)+\varphi(j-1),
\end{equation}
Eq.~\eqref{eq:dominant_equation} may be written compactly as
\begin{equation}
\Big[
-J\Delta_j
+ V_j
- \frac{i\gamma}{2}
+ \gamma|\varphi_0(j)|^2
\Big]\varphi_0(j)
=
E_{\mathrm{dom}}\varphi_0(j).
\end{equation}
For $\varepsilon\ll1$, the diagonal terms satisfy $|V_j|,\gamma \gg J$, so the operator is diagonally dominated. At $J=0$ the sites decouple, and the solution is localized on a single site $j_0$. Turning on small $J$ yields
\begin{equation}
|\varphi_0(j_0\pm1)|=\mathcal{O}(\varepsilon),
\qquad
|\varphi_0(j_0+n)|=\mathcal{O}(\varepsilon^n),
\end{equation}
and therefore exponential localization,
\begin{equation}
|\varphi_0(j)|\le C_0 e^{-|j-j_0|/\xi_0},
\qquad
\xi_0^{-1}=\ln(1/\varepsilon)=\mathcal{O}(1).
\end{equation}

% ============================================================
\subsection*{B. Fluctuation modes}
% ============================================================

We now linearize around the dominant solution by writing
\begin{equation}
\psi=\varphi_0+\delta\psi .
\end{equation}
Expanding to first order gives
\begin{equation}
|\psi|^2\psi
=
|\varphi_0|^2\varphi_0
+
2|\varphi_0|^2\delta\psi
+
\varphi_0^2\delta\psi^*
+
\mathcal{O}(\delta\psi^2).
\end{equation}

Subtracting the dominant equation yields the linearized fluctuation equation
\begin{align}
\Big[
{-}J\Delta_j
+ V_j
- \frac{i\gamma}{2}
+ 2\gamma|\varphi_0(j)|^2
- E_{\mathrm{dom}}
\Big]\varphi_\alpha(j)
=
\lambda_\alpha \varphi_\alpha(j),
\end{align}
which defines the fluctuation operator
\begin{equation}
M
=
-J\Delta
+ V
- \frac{i\gamma}{2}
+ 2\gamma|\varphi_0|^2
- E_{\mathrm{dom}} .
\end{equation}
Since diagonal dominance persists for $\varepsilon\ll1$, standard one-dimensional Combes--Thomas bounds~\cite{CombesThomas1973} imply exponential localization of all fluctuation modes,
\begin{equation}
|\varphi_\alpha(j)| \le C_\alpha e^{-|j-j_\alpha|/\xi_\alpha},
\qquad
\xi_\alpha=\mathcal{O}(1).
\end{equation}
Moreover, inside the localization core
\(
\mathcal C=\{j:\ |j-j_0|\lesssim\xi_0\},
\)
diagonal dominance implies additional suppression,
\begin{equation}
|\varphi_\alpha(j)|\le C'_\alpha\,\varepsilon,
\end{equation}
so fluctuations are parametrically small where the dominant mode has support.

Since each localized mode occupies $\mathcal{O}(1)$ sites and $\mathcal C$ has finite extent, only finitely many orthonormal modes can overlap significantly with $\mathcal C$. The effective local Hilbert-space dimension is therefore finite in the Zeno regime.

\section{Stratonovich--It\^o correction and deterministic backaction}
\label{app:ito_backaction}

In this appendix, we derive the deterministic It\^o backaction appearing in the linearized stochastic evolution by explicitly converting the Stratonovich equation to It\^o form~\cite{Pesce2013, reis2021relationstratonovichitointegrals}.

We consider the linear stochastic differential equation for the coefficients $C_\alpha$,
\begin{align}
    dC_\alpha
    &=
    \Big(
    - \sum_\beta M_{\alpha\beta} C_\beta
    + b_\alpha
    \Big) dt
    \nonumber\\
    &\quad
    +
    \sum_j
    \Big(
    U_{j\alpha}
    +
    \sum_\beta A_{j\alpha\beta} C_\beta
    \Big)\circ dW_j ,
    \label{eq:strat_sde_index}
\end{align}
where $M_{\alpha\beta}$ is a deterministic drift matrix, $b_\alpha$ is a constant vector, and $dW_j$ are independent Wiener increments satisfying
\begin{equation}
    dW_j dW_k = \delta_{jk}\, dt .
\end{equation}
The symbol $\circ\, dW_j$ denotes a Stratonovich stochastic integral, in contrast to the It\^o interpretation used for $dW_j$ without the circle.
The noise contains both an additive component $U_{j\alpha}$ and a multiplicative component proportional to $C_\beta$.

To convert Eq.~\eqref{eq:strat_sde_index} to It\^o form, we use the standard Stratonovich--It\^o conversion formula. For a Stratonovich equation
\begin{equation}
    dC_\alpha
    =
    f_\alpha(C)\,dt
    +
    \sum_j g_{j\alpha}(C)\circ dW_j ,
\end{equation}
the equivalent It\^o equation is
\begin{equation}
    dC_\alpha
    =
    \Big[
    f_\alpha(C)
    +
    \frac{1}{2}
    \sum_{j,\beta}
    \frac{\partial g_{j\alpha}}{\partial C_\beta}
    g_{j\beta}(C)
    \Big] dt
    +
    \sum_j g_{j\alpha}(C)\, dW_j .
\end{equation}In the present case, the noise amplitude is
\begin{equation}
    g_{j\alpha}(C)
    =
    U_{j\alpha}
    +
    \sum_\beta A_{j\alpha\beta} C_\beta ,
\end{equation}
and its Jacobian with respect to $C_\beta$ is therefore
\begin{equation}
    \frac{\partial g_{j\alpha}}{\partial C_\beta}
    =
    A_{j\alpha\beta}.
\end{equation}
Substituting into the It\^o correction term yields
\begin{align}
    \Delta f_\alpha
    &=
    \frac{1}{2}
    \sum_{j,\beta}
    A_{j\alpha\beta}
    \Big(
    U_{j\beta}
    +
    \sum_\gamma A_{j\beta\gamma} C_\gamma
    \Big)
    \nonumber\\
    &=
    \frac{1}{2}
    \sum_{j,\beta}
    A_{j\alpha\beta} U_{j\beta}
    +
    \frac{1}{2}
    \sum_{j,\beta,\gamma}
    A_{j\alpha\beta}
    A_{j\beta\gamma}
    C_\gamma .
\end{align}
The first term is independent of $C$ and can be absorbed into the constant drift vector $b_\alpha$. The second term is linear in $C$ and represents the deterministic It\^o backaction induced by the multiplicative noise. Retaining this contribution, the It\^o form of the stochastic equation becomes
\begin{align}
    dC_\alpha
    &=
    \Big(
    {-} \sum_\beta M_{\alpha\beta} C_\beta
    + b_\alpha
    \Big) dt
    \nonumber\\
    &\quad
    +
    \frac{1}{2}
    \sum_{j,\beta,\gamma}
    A_{j\alpha\beta}
    A_{j\beta\gamma}
    C_\gamma \, dt
    \nonumber\\
    &\quad
    +
    \sum_j
    \Big(
    U_{j\alpha}
    +
    \sum_\beta A_{j\alpha\beta} C_\beta
    \Big) dW_j .
    \label{eq:ito_final_index}
\end{align}
Eq.~\eqref{eq:ito_final_index} shows that the multiplicative noise produces an additional deterministic drift that renormalizes the linear drift matrix according to
\begin{equation}
    M_{\alpha\gamma}
    \;\longrightarrow\;
    M_{\alpha\gamma}
    -
    \frac{1}{2}
    \sum_{j,\beta}
    A_{j\alpha\beta}
    A_{j\beta\gamma}.
\end{equation}
This deterministic It\^o backaction arises solely from the multiplicative component of the noise and has a positive sign in the Stratonovich--It\^o conversion.

% ================== START APPENDIX ==================

% ----------------------------------------------------
\section{Diagonal cancellation in the projected QSD drift}
\label{app:diag_cancellation}

In this appendix, we give a detailed and explicit derivation of the cancellation of the leading $\mathcal{O}(\gamma)$ mode-dependent contributions in the linearized quantum state diffusion (QSD) dynamics after projection onto the fluctuation subspace orthogonal to the reference orbital $\varphi_0$. This cancellation plays a central role in establishing the $\mathcal{O}(\epsilon^2 \gamma)$ scaling of the effective damping rate in the Zeno regime $\gamma \gg J$.

Recall the It\^o correction term $\frac{1}{2}\sum_j A_j^2$ from Eq.~\eqref{eq:M_def_j}, associated with the multiplicative noise in the stochastic differential equation.
By the definition of matrix multiplication on the fluctuation subspace,
\begin{equation}
    (A_j^2)_{\alpha\beta}
    =
    \sum_{\eta \ge 1}
    (A_j)_{\alpha\eta} (A_j)_{\eta\beta}.
\end{equation}
Substituting Eq.~\eqref{eq:Aj} yields
\begin{align}
    \frac{1}{2}\sum_j (A_j^2)_{\alpha\beta}
    &=
    \frac{\gamma}{2}
    \sum_j \sum_{\eta \ge 1}
    \bigl[
    \varphi_\alpha^*(j)\varphi_\eta(j)
    -
    \delta_{\alpha\eta} a_j
    \bigr]
    \nonumber\\
    &\hspace{2em}\times
    \bigl[
    \varphi_\eta^*(j)\varphi_\beta(j)
    -
    \delta_{\eta\beta} a_j
    \bigr].
    \label{eq:A2_expand_app}
\end{align}

Using the restricted completeness relation
\begin{equation}
    \sum_{\eta \ge 1} \varphi_\eta(j)\varphi_\eta^*(j)
    =
    1 - |\varphi_0(j)|^2
    =
    1 - a_j ,
    \label{eq:restricted_completeness_app}
\end{equation}
and collecting all terms, one obtains
\begin{equation}
    \frac{1}{2}\sum_j (A_j^2)_{\alpha\beta}
    =
    \frac{\gamma}{2}
    \sum_j
    \bigl[
    (1 - 3 a_j)\varphi_\alpha^*(j)\varphi_\beta(j)
    +
    a_j^2 \delta_{\alpha\beta}
    \bigr].
    \label{eq:A2_result_app}
\end{equation}

We now combine the deterministic backaction term $Q_{\alpha\beta}$ [Eq.~\eqref{eq:Q_def_j}] with the It\^o correction [Eq.~\eqref{eq:A2_result_app}]. A direct algebraic addition shows that all leading $\mathcal{O}(\gamma)$ mode-dependent contributions cancel exactly, leaving
\begin{equation}
    Q_{\alpha\beta}
    +
    \frac{1}{2}\sum_j (A_j^2)_{\alpha\beta}
    =
    -\frac{\gamma}{2}
    \sum_j
    a_j \,
    \varphi_\alpha^*(j)\varphi_\beta(j).
    \label{eq:final_cancellation_app}
\end{equation}
This identity holds exactly and does not rely on any approximation. The cancellation proceeds through a precise pairwise elimination of terms: the $\sum_j \varphi_\alpha^*(j)\varphi_\beta(j)$ contributions without $a_j$ cancel between $Q$ and the It\^o correction; the terms proportional to $a_j \varphi_\alpha^*(j)\varphi_\beta(j)$ cancel in the same manner; and the diagonal terms proportional to $a_j^2 \delta_{\alpha\beta}$ vanish identically.

The remaining term in Eq.~\eqref{eq:final_cancellation_app} still carries a prefactor $\gamma$, but it is weighted by the localized density $a_j$. In the quantum Zeno regime $\gamma \gg J$, the reference mode $\varphi_0$ is exponentially localized around a site $j_0$ with localization length $\xi = \mathcal{O}(1)$. Consequently, $a_j$ is of order unity only within the localization core $|j-j_0| \lesssim \xi$ and is exponentially small outside this region. The fluctuation modes $\varphi_\alpha$ ($\alpha \ge 1$), being orthogonal to $\varphi_0$, satisfy within the core the amplitude bound
\begin{equation}
    |\varphi_\alpha(j)| \lesssim \varepsilon ,
\end{equation}
which follows from balancing the hopping term of order $J$ against the measurement-induced term of order $\gamma$ in the linearized eigenvalue equation.

As a consequence,
\begin{equation}
    a_j \, \varphi_\alpha^*(j)\varphi_\beta(j)
    =
    \mathcal{O}(\varepsilon^2)
\end{equation}
on the sites that contribute appreciably, and negligible elsewhere. Since only $\mathcal{O}(1)$ sites lie within the localization core, one finds
\begin{equation}
    \sum_j a_j\,\varphi_\alpha^*(j)\varphi_\beta(j)
    =
    \mathcal{O}(\varepsilon^2).
\end{equation}
Multiplying by the prefactor $-\gamma/2$ in Eq.~\eqref{eq:final_cancellation_app} gives
\begin{equation}
    Q_{\alpha\beta}
    +
    \frac{1}{2}\sum_j (A_j^2)_{\alpha\beta}
    =
    \mathcal{O}(\varepsilon^2 \gamma).
\end{equation}
Thus, after projection onto the fluctuation subspace, all $\mathcal{O}(\gamma)$ mode-dependent contributions to the linearized drift are eliminated. What remains is a mode-independent constant, which may be absorbed into a global phase, together with mode-dependent terms suppressed by $\varepsilon^2 \gamma$. The real parts of the eigenvalues of the total drift matrix $M$ are therefore of order $\varepsilon^2 \gamma$, providing a finite spectral gap that guarantees exponential relaxation of fluctuations and validates the linearized description. This cancellation mechanism underlies the emergence of an effective non-Hermitian Hamiltonian
\begin{equation}
    H_{\mathrm{eff}} = h - i\frac{\gamma}{2} I
\end{equation}
governing the dynamics in the Zeno limit.

% ----------------------------------------------------
\section{Uniform bound on fluctuation amplitudes}
\label{app:fluct_estimate}

In this appendix, we establish a uniform-in-time bound on the second moment of the fluctuation amplitudes $c_\alpha(t)$ starting from the linearized stochastic equations~\eqref{eq:lin_sde_c}. For $\alpha\neq0$ the dynamics has the schematic form
\begin{equation}
    dc_\alpha
    =
    -\mu_\alpha c_\alpha\,dt
    +
    \sum_j \bigl(A_{\alpha j} c_\alpha + u_{\alpha j}\bigr)\, dW_j ,
\end{equation}
where $dW_j$ are independent Wiener increments. In the quantum Zeno regime, the projected generator has a strictly positive real part, and one has
\begin{equation}
    \mu_\alpha \gtrsim \varepsilon^2 \gamma,
    \qquad
    \varepsilon = \frac{J}{\sqrt{\lambda^2+(\gamma/2)^2}}\ll1.
\end{equation}
The precise prefactor is unimportant; only the positivity of $\mu_\alpha$ is required.

Applying It\^o's formula to $|c_\alpha|^2$ gives
\begin{equation}
    d|c_\alpha|^2
    =
    -2\mu_\alpha |c_\alpha|^2 dt
    +
    \sum_j |A_{\alpha j} c_\alpha + u_{\alpha j}|^2 dt
    +
    dM_t ,
\end{equation}
where $M_t$ is a martingale with $\mathbb{E}[dM_t]=0$. Taking expectations yields
\begin{equation}
    \frac{d}{dt}\mathbb{E}|c_\alpha|^2
    =
    -2\mu_\alpha \mathbb{E}|c_\alpha|^2
    +
    \sum_j
    \mathbb{E}|A_{\alpha j} c_\alpha + u_{\alpha j}|^2 .
\end{equation}
Expanding the square and using Young's inequality,
\begin{equation}
    |A_{\alpha j} c_\alpha + u_{\alpha j}|^2
    \le
    2 |A_{\alpha j}|^2 |c_\alpha|^2
    +
    2 |u_{\alpha j}|^2 ,
\end{equation}
we obtain the differential inequality
\begin{equation}
    \frac{d}{dt}\mathbb{E}|c_\alpha|^2
    \le
    -\bigl(2\mu_\alpha - 2\sum_j |A_{\alpha j}|^2\bigr)
    \mathbb{E}|c_\alpha|^2
    +
    2\sum_j |u_{\alpha j}|^2 .
\end{equation}
In the Zeno regime, the leading $\mathcal{O}(\gamma)$ diagonal contributions cancel in the projected generator, so both $\mu_\alpha$ and $\sum_j |A_{\alpha j}|^2$ scale as $\mathcal{O}(\varepsilon^2\gamma)$, and the effective damping coefficient remains strictly positive and of order $\varepsilon^2\gamma$.

The additive source terms are
\begin{equation}
    u_{\alpha j} = \sqrt{\gamma}\,\phi_\alpha^*(j)\phi_0(j).
\end{equation}
Since $|\phi_\alpha(j)| = \mathcal{O}(\varepsilon)$ within the localization core of $\phi_0$ and both modes are exponentially localized, one finds
\begin{equation}
    \sum_j |u_{\alpha j}|^2
    =
    \gamma \sum_j |\phi_\alpha(j)|^2 |\phi_0(j)|^2
    =
    \mathcal{O}(\varepsilon^2\gamma).
\end{equation}
Consequently,
\begin{equation}
    \frac{d}{dt}\mathbb{E}|c_\alpha|^2
    \le
    -a\,\mathbb{E}|c_\alpha|^2
    +
    b ,
\end{equation}
with $a,b=\mathcal{O}(\varepsilon^2\gamma)$ and $a>0$. Gr\"onwall's inequality then implies the uniform bound
\begin{equation}
    \sup_{t\ge0}\mathbb{E}|c_\alpha(t)|^2
    \le
    \frac{b}{a}
    =
    \mathcal{O}(1).
\end{equation}
Since only $\mathcal{O}(1)$ fluctuation modes have appreciable overlap with the localization core of the dominant mode, it follows that
\begin{equation}
    \sup_{t\ge0}\mathbb{E}\|c(t)\|^2
    =
    \sup_{t\ge0}\sum_{\alpha\neq0}\mathbb{E}|c_\alpha(t)|^2
    =
    \mathcal{O}(1).
\end{equation}

The resulting second-moment bound is parametrically consistent with the stationary Ornstein--Uhlenbeck variance~\eqref{OU_balance}, reflecting the identical scaling of measurement-induced noise and dissipation in the Zeno regime.

\section{Numerical check of the bound on the fluctuation field}
\label{app:bound_check}

In this appendix, we show that the bound in Eq.~\eqref{eq:delta_psi_scaling} is satisfied by the QSD numerics. First, we define $\delta$ as the probability of finding a particle outside the localization center,
\begin{equation}
  \delta^2 = \sum_{j\neq j_0} |\psi_j|^2 = 1 - |\psi_{j_0}|^2,
\end{equation}
where $j_0$ is the location of the localization center. Next, we decompose the wave function into a dominant localized mode $\phi_0$ and a fluctuation field $\delta\psi$. We assume $\phi_0$ is exponentially localized around site $j_0$, carrying O(1) weight in the localization core and exponentially small tails. Then, since for $j\neq j_0$, $|\psi_j|^2 \approx |\delta \psi_j|^2$, we notice that $\delta$ provides an estimate of the norm of the fluctuation field $\delta\psi$ to leading order,
\begin{equation}
  \delta^2 \approx \sum_{j \neq j_0} |\delta \psi_j|^2 = ||\delta \psi||^2.
\end{equation}
Using the bound from Eq.~\eqref{eq:delta_psi_scaling}, and since only $O(1)$ sites contribute significantly within the localization core, this implies
\begin{equation}
  \delta \lesssim \mathcal{O} (\varepsilon) = \mathcal{O}\! \left( \frac{J}{\sqrt{\lambda^2 + (\gamma/2)^2 }} \right).
  \label{eq:delta_bound}
\end{equation}
Note that this bound assumes that $\phi_0$ has only significant support on site $j_0$, which may not be true for certain parameter regimes, but is correct when a Zeno limit is taken.

\begin{figure}
    \centering
    \includegraphics[width=\columnwidth]{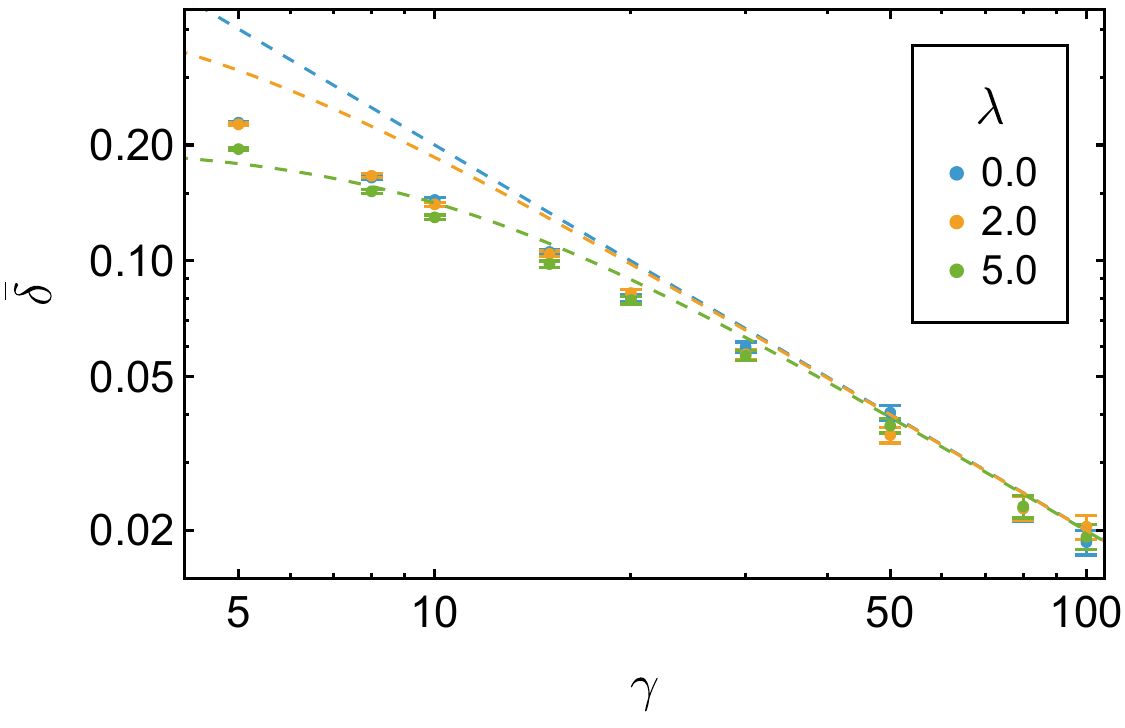}
    \caption{Comparison between the QSD numerics (markers) and the bound in Eq.~\eqref{eq:delta_bound} (dashed lines).}
    \label{fig:delta}
\end{figure}

We verify this bound numerically using QSD evolution, with parameters detailed in Sec.~\ref{sec:results}. For each unscrambled single-particle wave function, we calculate $\delta^2$, which is then averaged over 100 trajectories and all orbitals. The resulting $\bar{\delta} = \sqrt{\langle \delta^2 \rangle}$ as a function of $\gamma$ for different values of $\lambda$ is plotted in Fig.~\ref{fig:delta}. The numerical results are consistently below $\bar\delta = \varepsilon$ (shown in dashed lines), confirming the bound in Eq.~\eqref{eq:delta_psi_scaling}.

\section{Transfer-matrix derivation of asymptotic Lyapunov exponents}
\label{app:transfer_matrix}

In this appendix, we present a systematic transfer-matrix derivation of the Lyapunov exponent $\kappa(\lambda,\gamma)$ for the monitored Aubry--Andr\'e--Harper (AAH) model. We first introduce the general transfer-matrix formalism at vanishing real energy and then analyze several asymptotic parameter regimes relevant to the localization properties discussed in the main text.

%-------------------------------------------------
\subsection{General transfer-matrix framework}
\label{app:general}

%\subsubsection{Single-particle recurrence}

We consider the single-particle tight-binding equation at zero real energy, $ E_{\mathrm{dom}} = 0$,
\begin{equation}
    - J(\psi_{j+1}+\psi_{j-1})
    + \left[\lambda\cos(2\pi\alpha j+\theta)-i\frac{\gamma}{2}\right]\psi_j = 0 .
    \label{eq:app_recurrence}
\end{equation}
Solving for $\psi_{j+1}$ and introducing the two-component state vector $\Psi_j=(\psi_j,\psi_{j-1})^{\mathsf T}$, Eq.~\eqref{eq:app_recurrence} can be cast into the one-step transfer-matrix form
\begin{equation}
    \Psi_{j+1} = T_j \Psi_j .
    \label{eq:app_transfer_relation}
\end{equation}
%
%\subsubsection{Transfer matrix}
%
The site-dependent transfer matrix is given by
\begin{equation}
    T_j =
    \begin{pmatrix}
    a(\phi_j) & -1 \\
    1 & 0
    \end{pmatrix},
    \qquad
    a(\phi) \equiv \frac{\lambda}{J}\cos\phi - i\frac{\gamma}{2J},
    \label{eq:app_Tj}
\end{equation}
where $\phi_j = 2\pi\alpha j + \theta$ denotes the quasiperiodic phase.
%\subsubsection{Lyapunov exponent}
For irrational $\alpha$, the sequence $\{\phi_j\}$ is ergodic.

Let us now use the definition of the Lyapunov exponent from Eq.~\eqref{eq:kappa_definition}, and since $\{\phi_j\}$ is ergodic, we can insert an average over $\phi$ before taking the limit,
\begin{equation}
    \kappa = \lim_{n\to\infty} \frac{1}{n} \int \frac{d\phi}{2\pi} \log \Big\| \prod_{j=1}^{n} T_j \Big\|.
\end{equation}
Next we use matrix norm submultiplicativity ($\|AB\| < \|A\| \|B\|$) to bound the expression from above, 
\begin{align}
    \kappa &< \lim_{n\to\infty} \frac{1}{n} \int \frac{d\phi}{2\pi} \sum_{j=1}^{n} \log \| T_j \| \\
    &= \lim_{n\to\infty} \frac{1}{n} \sum_{j=1}^{n} \int \frac{d\phi}{2\pi}\log \| T_j \|.
\end{align}
Integration on the right will give the same result independently of $j$, hence we can drop the sum, as well as the limit. When the system is deep in the Zeno regime, there is a large term in the transfer matrix (the $\gamma$ term) and multiplying $T_j$ has an accumulating effect on the norm (rather than, e.g., $\|\prod T_j\|$ being bounded by some finite value). Hence, $\|\prod T_j\| \approx \prod\| T_j\|$ and the bound will be very close to the value of the Lyapunov exponent $\kappa$, and can be used as its estimate,
%The Lyapunov exponent governing spatial localization is therefore obtained by phase averaging,
\begin{equation}
    \kappa(\lambda,\gamma)
    \approx \int_0^{2\pi}\frac{d\phi}{2\pi}\,\ln\|T(\phi)\|
    = \int_0^{2\pi}\frac{d\phi}{2\pi}\,\ln s_+(\phi),
    \label{eq:app_kappa_def}
\end{equation}
where $s_+(\phi)$ is the largest singular value of $T(\phi)$. Defining $\nu_+(\phi)=s_+^2(\phi)$, a straightforward evaluation of $T^\dagger T$ ~\cite{PhysRevB.110.134203,PhysRevB.104.104203} yields
\begin{equation}
    \nu_+(\phi)
    = \frac{|a(\phi)|^2+2+\sqrt{(|a(\phi)|^2+2)^2-4}}{2},
    \label{eq:app_nu_plus}
\end{equation}
with
\begin{equation}
    |a(\phi)|^2
    = \left(\frac{\lambda}{J}\cos\phi\right)^2
    + \left(\frac{\gamma}{2J}\right)^2 .
    \label{eq:app_a_squared}
\end{equation}
Depending on whether $|a(\phi)|$ is large or small compared to unity, distinct asymptotic behaviors of $\kappa(\lambda,\gamma)$ emerge, which we analyze below.

%-------------------------------------------------
\subsection{Measurement-dominated regime}
\label{app:zeno}

We first consider the pure measurement limit in the absence of a quasiperiodic potential,
\begin{equation}
    \lambda = 0, \qquad \gamma > 0 .
\end{equation}
In this case, the recurrence relation reduces to
\begin{equation}
    - J(\psi_{j+1}+\psi_{j-1}) - i\frac{\gamma}{2}\psi_j = 0 .
\end{equation}
The corresponding transfer matrix is site-independent and reads
\begin{equation}
    T =
    \begin{pmatrix}
    -i\delta & -1 \\
    1 & 0
    \end{pmatrix},
    \qquad
    \delta \equiv \frac{\gamma}{2J}.
\end{equation}
The largest eigenvalue of $T^\dagger T$ is
\begin{equation}
    \nu_+ = \frac{\delta^2+2+\delta\sqrt{\delta^2+4}}{2}.
\end{equation}
Introducing the parametrization $\sinh y = \delta/2$, the operator norm becomes $\|T\| = e^{y}$, leading to the exact Lyapunov exponent
\begin{equation}
    \kappa(0,\gamma)
    \approx \operatorname{arsinh}\!\left(\frac{\gamma}{4J}\right).
    \label{eq:app_kappa_zeno_exact}
\end{equation}
\textit{This result shows that continuous measurement alone induces exponential spatial localization, even in the absence of disorder or quasiperiodic modulation.}

%-------------------------------------------------
% \subsection{Regime II: Measurement-dominated localization with weak quasiperiodicity}
% \label{app:meas_dominated}

We next consider strong measurement in the presence of a weak quasiperiodic potential,
\begin{equation}
    \frac{\gamma}{2J} \gg 1,
    \qquad
    \frac{\lambda}{\gamma} \ll 1 .
\end{equation}
In this regime $|a(\phi)| \gg 1$ for typical phases, allowing for a controlled asymptotic expansion.

To leading order,
\begin{equation}
    \ln s_+(\phi)
    = \ln|a(\phi)| + |a(\phi)|^{-2}
    + \mathcal{O}(|a|^{-4}),
\end{equation}
and the Lyapunov exponent decomposes as
\begin{equation}
    \kappa(\lambda,\gamma)
    \approx I_1 + I_2 + \mathcal{O}(\gamma^{-4}),
\end{equation}
where
\begin{equation}
    I_1 = \int_0^{2\pi}\frac{d\phi}{2\pi}\ln|a(\phi)|,
    \qquad
    I_2 = \int_0^{2\pi}\frac{d\phi}{2\pi}|a(\phi)|^{-2}.
\end{equation}
Evaluating these integrals yields
\begin{equation}
    \kappa(\lambda,\gamma)
    \approx \operatorname{arsinh}\!\left(\frac{\gamma}{4J}\right)
    + \frac{\lambda^2}{\gamma^2}
    + \mathcal{O}\!\left(\frac{J^2\lambda^2}{\gamma^4},\frac{\lambda^4}{\gamma^4}\right).
    \label{eq:app_kappa_measdom}
\end{equation}

\textit{The dominant Zeno-induced localization is weakly renormalized by the quasiperiodic potential, leading to a perturbative enhancement of the Lyapunov exponent.}

%-------------------------------------------------
\subsection{Strong quasiperiodic potential and strong measurement regime}
\label{app:strong_strong}

Finally, we consider the regime in which both the quasiperiodic potential and the measurement
strength dominate the hopping amplitude,
\begin{equation}
    \min(\lambda,\gamma) \gg J .
\end{equation}
In this limit the transfer-matrix norm satisfies $\|T(\phi)\| \approx |a(\phi)|$.

The Lyapunov exponent then takes the form
\begin{equation}
    \kappa(\lambda,\gamma)
    \approx \int_0^{2\pi}\frac{d\phi}{2\pi}
    \ln\!\left(
    \frac{\sqrt{(\lambda\cos\phi)^2 + (\gamma/2)^2}}{J}
    \right),
\end{equation}
which evaluates to
\begin{equation}
    \kappa(\lambda,\gamma)
    \approx \ln\!\left|\frac{\lambda}{2J}\right|
    + \operatorname{arsinh}\!\left(\frac{\gamma}{2\lambda}\right)
    + \mathcal{O}\!\left(\frac{J\varepsilon}{\gamma}\right).
    \label{eq:app_kappa_strong}
\end{equation}

\textit{In this regime, quasiperiodic and measurement-induced localization mechanisms act cooperatively, giving rise to additive contributions to the Lyapunov exponent.}

\bibliography{ref}

@article{PhysRevB.100.134306,
  title = {Measurement-driven entanglement transition in hybrid quantum circuits},
  author = {Li, Yaodong and Chen, Xiao and Fisher, Matthew P. A.},
  journal = {Phys. Rev. B},
  volume = {100},
  issue = {13},
  pages = {134306},
  numpages = {26},
  year = {2019},
  month = {Oct},
  publisher = {American Physical Society},
  doi = {10.1103/PhysRevB.100.134306},
  url = {https://link.aps.org/doi/10.1103/PhysRevB.100.134306}
}

@article{PhysRevB.98.205136,
  title = {{Quantum Zeno effect and the many-body entanglement transition}},
  author = {Li, Yaodong and Chen, Xiao and Fisher, Matthew P. A.},
  journal = {Phys. Rev. B},
  volume = {98},
  issue = {20},
  pages = {205136},
  numpages = {9},
  year = {2018},
  month = {Nov},
  publisher = {American Physical Society},
  doi = {10.1103/PhysRevB.98.205136},
  url = {https://link.aps.org/doi/10.1103/PhysRevB.98.205136}
}

@article{Biella2021manybodyquantumzeno,
  doi = {10.22331/q-2021-08-19-528},
  url = {https://doi.org/10.22331/q-2021-08-19-528},
  title = {Many-body quantum {Z}eno effect and measurement-induced subradiance transition},
  author = {Biella, Alberto and Schir{\'{o}}, Marco},
  journal = {{Quantum}},
  issn = {2521-327X},
  publisher = {{Verein zur F{\"{o}}rderung des Open Access Publizierens in den Quantenwissenschaften}},
  volume = {5},
  pages = {528},
  month = aug,
  year = {2021}
}

@article{PhysRevX.9.031009,
  title = {Measurement-Induced Phase Transitions in the Dynamics of Entanglement},
  author = {Skinner, Brian and Ruhman, Jonathan and Nahum, Adam},
  journal = {Phys. Rev. X},
  volume = {9},
  issue = {3},
  pages = {031009},
  numpages = {21},
  year = {2019},
  month = {Jul},
  publisher = {American Physical Society},
  doi = {10.1103/PhysRevX.9.031009},
  url = {https://link.aps.org/doi/10.1103/PhysRevX.9.031009}
}

@article{PhysRevB.108.165126,
  title = {Disordered monitored free fermions},
  author = {Szyniszewski, Marcin and Lunt, Oliver and Pal, Arijeet},
  journal = {Phys. Rev. B},
  volume = {108},
  issue = {16},
  pages = {165126},
  numpages = {12},
  year = {2023},
  month = {Oct},
  publisher = {American Physical Society},
  doi = {10.1103/PhysRevB.108.165126},
  url = {https://link.aps.org/doi/10.1103/PhysRevB.108.165126}
}

@article{PRXQuantum.4.040332,
  title = {Ultrafast Entanglement Dynamics in Monitored Quantum Circuits},
  author = {Sang, Shengqi and Li, Zhi and Hsieh, Timothy H. and Yoshida, Beni},
  journal = {PRX Quantum},
  volume = {4},
  issue = {4},
  pages = {040332},
  numpages = {20},
  year = {2023},
  month = {Nov},
  publisher = {American Physical Society},
  doi = {10.1103/PRXQuantum.4.040332},
  url = {https://link.aps.org/doi/10.1103/PRXQuantum.4.040332}
}

@article{PhysRevResearch.7.023082,
  title = {Measurement-induced phase transitions in monitored infinite-range interacting systems},
  author = {Delmonte, Anna and Li, Zejian and Passarelli, Gianluca and Song, Eric Yilun and Barberena, Diego and Rey, Ana Maria and Fazio, Rosario},
  journal = {Phys. Rev. Res.},
  volume = {7},
  issue = {2},
  pages = {023082},
  numpages = {27},
  year = {2025},
  month = {Apr},
  publisher = {American Physical Society},
  doi = {10.1103/PhysRevResearch.7.023082},
  url = {https://link.aps.org/doi/10.1103/PhysRevResearch.7.023082}
}

@article{PhysRevB.100.064204,
  title = {Entanglement transition from variable-strength weak measurements},
  author = {Szyniszewski, M. and Romito, A. and Schomerus, H.},
  journal = {Phys. Rev. B},
  volume = {100},
  issue = {6},
  pages = {064204},
  numpages = {8},
  year = {2019},
  month = {Aug},
  publisher = {American Physical Society},
  doi = {10.1103/PhysRevB.100.064204},
  url = {https://link.aps.org/doi/10.1103/PhysRevB.100.064204}
}

@article{PhysRevLett.125.030505,
  title = {Quantum Error Correction in Scrambling Dynamics and Measurement-Induced Phase Transition},
  author = {Choi, Soonwon and Bao, Yimu and Qi, Xiao-Liang and Altman, Ehud},
  journal = {Phys. Rev. Lett.},
  volume = {125},
  issue = {3},
  pages = {030505},
  numpages = {6},
  year = {2020},
  month = {Jul},
  publisher = {American Physical Society},
  doi = {10.1103/PhysRevLett.125.030505},
  url = {https://link.aps.org/doi/10.1103/PhysRevLett.125.030505}
}

@Article{10.21468/SciPostPhysCore.5.2.023,
	title={{Measurement-induced criticality in extended and long-range unitary circuits}},
	author={Shraddha Sharma and Xhek Turkeshi and Rosario Fazio and Marcello Dalmonte},
	journal={SciPost Phys. Core},
	volume={5},
	pages={023},
	year={2022},
	publisher={SciPost},
	doi={10.21468/SciPostPhysCore.5.2.023},
	url={https://scipost.org/10.21468/SciPostPhysCore.5.2.023},
}

@article{PhysRevX.12.041002,
  title = {{Entanglement and charge-sharpening transitions in $U(1)$ symmetric monitored quantum circuits}},
  author = {Agrawal, Utkarsh and Zabalo, Aidan and Chen, Kun and Wilson, Justin H. and Potter, Andrew C. and Pixley, J. H. and Gopalakrishnan, Sarang and Vasseur, Romain},
  journal = {Phys. Rev. X},
  volume = {12},
  issue = {4},
  pages = {041002},
  numpages = {29},
  year = {2022},
  month = {Oct},
  publisher = {American Physical Society},
  doi = {10.1103/PhysRevX.12.041002},
  url = {https://link.aps.org/doi/10.1103/PhysRevX.12.041002}
}

@article{PhysRevLett.129.120604,
  title = {Field Theory of Charge Sharpening in Symmetric Monitored Quantum Circuits},
  author = {Barratt, Fergus and Agrawal, Utkarsh and Gopalakrishnan, Sarang and Huse, David A. and Vasseur, Romain and Potter, Andrew C.},
  journal = {Phys. Rev. Lett.},
  volume = {129},
  issue = {12},
  pages = {120604},
  numpages = {7},
  year = {2022},
  month = {Sep},
  publisher = {American Physical Society},
  doi = {10.1103/PhysRevLett.129.120604},
  url = {https://link.aps.org/doi/10.1103/PhysRevLett.129.120604}
}

@Article{10.21468/SciPostPhys.15.6.250,
	title={{Coherence requirements for quantum communication from hybrid circuit dynamics}},
	author={Shane P. Kelly and Ulrich Poschinger and Ferdinand Schmidt-Kaler and Matthew P. A. Fisher and Jamir Marino},
	journal={SciPost Phys.},
	volume={15},
	pages={250},
	year={2023},
	publisher={SciPost},
	doi={10.21468/SciPostPhys.15.6.250},
	url={https://scipost.org/10.21468/SciPostPhys.15.6.250},
}

@Article{10.21468/SciPostPhys.7.2.024,
	title={{Entanglement in a fermion chain under continuous monitoring}},
	author={Xiangyu Cao and Antoine Tilloy and Andrea De Luca},
	journal={SciPost Phys.},
	volume={7},
	pages={024},
	year={2019},
	publisher={SciPost},
	doi={10.21468/SciPostPhys.7.2.024},
	url={https://scipost.org/10.21468/SciPostPhys.7.2.024},
}

@article{PhysRevResearch.2.033017,
  title = {Emergent conformal symmetry in nonunitary random dynamics of free fermions},
  author = {Chen, Xiao and Li, Yaodong and Fisher, Matthew P. A. and Lucas, Andrew},
  journal = {Phys. Rev. Res.},
  volume = {2},
  issue = {3},
  pages = {033017},
  numpages = {15},
  year = {2020},
  month = {Jul},
  publisher = {American Physical Society},
  doi = {10.1103/PhysRevResearch.2.033017},
  url = {https://link.aps.org/doi/10.1103/PhysRevResearch.2.033017}
}

@article{PhysRevB.103.174303,
  title = {Quantum criticality in the nonunitary dynamics of $(2+1)$-dimensional free fermions},
  author = {Tang, Qicheng and Chen, Xiao and Zhu, W.},
  journal = {Phys. Rev. B},
  volume = {103},
  issue = {17},
  pages = {174303},
  numpages = {23},
  year = {2021},
  month = {May},
  publisher = {American Physical Society},
  doi = {10.1103/PhysRevB.103.174303},
  url = {https://link.aps.org/doi/10.1103/PhysRevB.103.174303}
}

@article{PhysRevB.105.094303,
  title = {Growth of entanglement entropy under local projective measurements},
  author = {Coppola, Michele and Tirrito, Emanuele and Karevski, Dragi and Collura, Mario},
  journal = {Phys. Rev. B},
  volume = {105},
  issue = {9},
  pages = {094303},
  numpages = {9},
  year = {2022},
  month = {Mar},
  publisher = {American Physical Society},
  doi = {10.1103/PhysRevB.105.094303},
  url = {https://link.aps.org/doi/10.1103/PhysRevB.105.094303}
}

@article{PhysRevResearch.4.033001,
  title = {Monitored open fermion dynamics: Exploring the interplay of measurement, decoherence, and free Hamiltonian evolution},
  author = {Ladewig, B. and Diehl, S. and Buchhold, M.},
  journal = {Phys. Rev. Res.},
  volume = {4},
  issue = {3},
  pages = {033001},
  numpages = {32},
  year = {2022},
  month = {Jul},
  publisher = {American Physical Society},
  doi = {10.1103/PhysRevResearch.4.033001},
  url = {https://link.aps.org/doi/10.1103/PhysRevResearch.4.033001}
}

@Article{10.21468/SciPostPhys.14.5.138,
	title={{Volume-to-area law entanglement transition in a non-Hermitian free fermionic chain}},
	author={Youenn Le Gal and Xhek Turkeshi and Marco Schirò},
	journal={SciPost Phys.},
	volume={14},
	pages={138},
	year={2023},
	publisher={SciPost},
	doi={10.21468/SciPostPhys.14.5.138},
	url={https://scipost.org/10.21468/SciPostPhys.14.5.138},
}

@article{PhysRevB.106.L220304,
  title = {Entangled multiplets and spreading of quantum correlations in a continuously monitored tight-binding chain},
  author = {Carollo, Federico and Alba, Vincenzo},
  journal = {Phys. Rev. B},
  volume = {106},
  issue = {22},
  pages = {L220304},
  numpages = {7},
  year = {2022},
  month = {Dec},
  publisher = {American Physical Society},
  doi = {10.1103/PhysRevB.106.L220304},
  url = {https://link.aps.org/doi/10.1103/PhysRevB.106.L220304}
}

@article{PhysRevResearch.5.033174,
  title = {Keldysh nonlinear sigma model for a free-fermion gas under continuous measurements},
  author = {Yang, Qinghong and Zuo, Yi and Liu, Dong E.},
  journal = {Phys. Rev. Res.},
  volume = {5},
  issue = {3},
  pages = {033174},
  numpages = {12},
  year = {2023},
  month = {Sep},
  publisher = {American Physical Society},
  doi = {10.1103/PhysRevResearch.5.033174},
  url = {https://link.aps.org/doi/10.1103/PhysRevResearch.5.033174}
}

@article{PhysRevX.11.041004,
  title = {Effective Theory for the Measurement-Induced Phase Transition of Dirac Fermions},
  author = {Buchhold, M. and Minoguchi, Y. and Altland, A. and Diehl, S.},
  journal = {Phys. Rev. X},
  volume = {11},
  issue = {4},
  pages = {041004},
  numpages = {35},
  year = {2021},
  month = {Oct},
  publisher = {American Physical Society},
  doi = {10.1103/PhysRevX.11.041004},
  url = {https://link.aps.org/doi/10.1103/PhysRevX.11.041004}
}

@article{PhysRevLett.126.123604,
  title = {Entanglement Entropy Scaling Transition under Competing Monitoring Protocols},
  author = {Van Regemortel, Mathias and Cian, Ze-Pei and Seif, Alireza and Dehghani, Hossein and Hafezi, Mohammad},
  journal = {Phys. Rev. Lett.},
  volume = {126},
  issue = {12},
  pages = {123604},
  numpages = {6},
  year = {2021},
  month = {Mar},
  publisher = {American Physical Society},
  doi = {10.1103/PhysRevLett.126.123604},
  url = {https://link.aps.org/doi/10.1103/PhysRevLett.126.123604}
}

@article{PhysRevB.106.024304,
  title = {Enhanced entanglement negativity in boundary-driven monitored fermionic chains},
  author = {Turkeshi, Xhek and Piroli, Lorenzo and Schir\'o, Marco},
  journal = {Phys. Rev. B},
  volume = {106},
  issue = {2},
  pages = {024304},
  numpages = {11},
  year = {2022},
  month = {Jul},
  publisher = {American Physical Society},
  doi = {10.1103/PhysRevB.106.024304},
  url = {https://link.aps.org/doi/10.1103/PhysRevB.106.024304}
}

@article{PhysRevX.13.041045,
  title = {Nonlinear Sigma Models for Monitored Dynamics of Free Fermions},
  author = {Fava, Michele and Piroli, Lorenzo and Swann, Tobias and Bernard, Denis and Nahum, Adam},
  journal = {Phys. Rev. X},
  volume = {13},
  issue = {4},
  pages = {041045},
  numpages = {33},
  year = {2023},
  month = {Dec},
  publisher = {American Physical Society},
  doi = {10.1103/PhysRevX.13.041045},
  url = {https://link.aps.org/doi/10.1103/PhysRevX.13.041045}
}

@article{PhysRevB.105.064305,
  title = {{Entanglement transitions in the quantum Ising chain: A comparison between different unravelings of the same Lindbladian}},
  author = {Piccitto, Giulia and Russomanno, Angelo and Rossini, Davide},
  journal = {Phys. Rev. B},
  volume = {105},
  issue = {6},
  pages = {064305},
  numpages = {14},
  year = {2022},
  month = {Feb},
  publisher = {American Physical Society},
  doi = {10.1103/PhysRevB.105.064305},
  url = {https://link.aps.org/doi/10.1103/PhysRevB.105.064305}
}

@Article{10.21468/SciPostPhysCore.6.4.078,
	title={{Entanglement dynamics with string measurement operators}},
	author={Giulia Piccitto and Angelo Russomanno and Davide Rossini},
	journal={SciPost Phys. Core},
	volume={6},
	pages={078},
	year={2023},
	publisher={SciPost},
	doi={10.21468/SciPostPhysCore.6.4.078},
	url={https://scipost.org/10.21468/SciPostPhysCore.6.4.078},
}

@article{PhysRevB.108.104313,
  title = {Entanglement transitions and quantum bifurcations under continuous long-range monitoring},
  author = {Russomanno, Angelo and Piccitto, Giulia and Rossini, Davide},
  journal = {Phys. Rev. B},
  volume = {108},
  issue = {10},
  pages = {104313},
  numpages = {11},
  year = {2023},
  month = {Sep},
  publisher = {American Physical Society},
  doi = {10.1103/PhysRevB.108.104313},
  url = {https://link.aps.org/doi/10.1103/PhysRevB.108.104313}
}

@Article{10.21468/SciPostPhys.14.3.031,
	title={{Topological transitions in weakly monitored free fermions}},
	author={Graham Kells and Dganit Meidan and Alessandro Romito},
	journal={SciPost Phys.},
	volume={14},
	pages={031},
	year={2023},
	publisher={SciPost},
	doi={10.21468/SciPostPhys.14.3.031},
	url={https://scipost.org/10.21468/SciPostPhys.14.3.031},
}

@article{PhysRevB.108.L020306,
  title = {Purification timescales in monitored fermions},
  author = {L\'oio, Hugo and De Luca, Andrea and De Nardis, Jacopo and Turkeshi, Xhek},
  journal = {Phys. Rev. B},
  volume = {108},
  issue = {2},
  pages = {L020306},
  numpages = {7},
  year = {2023},
  month = {Jul},
  publisher = {American Physical Society},
  doi = {10.1103/PhysRevB.108.L020306},
  url = {https://link.aps.org/doi/10.1103/PhysRevB.108.L020306}
}

@article{PhysRevX.13.041046,
  title = {Theory of Free Fermions under Random Projective Measurements},
  author = {Poboiko, Igor and P\"opperl, Paul and Gornyi, Igor V. and Mirlin, Alexander D.},
  journal = {Phys. Rev. X},
  volume = {13},
  issue = {4},
  pages = {041046},
  numpages = {26},
  year = {2023},
  month = {Dec},
  publisher = {American Physical Society},
  doi = {10.1103/PhysRevX.13.041046},
  url = {https://link.aps.org/doi/10.1103/PhysRevX.13.041046}
}

@article{Noel2022,
	author = {Noel, Crystal and Niroula, Pradeep and Zhu, Daiwei and Risinger, Andrew and Egan, Laird and Biswas, Debopriyo and Cetina, Marko and Gorshkov, Alexey V. and Gullans, Michael J. and Huse, David A. and others},
	title = {{Measurement-induced quantum phases realized in a trapped-ion quantum computer}},
	journal = {Nat. Phys.},
	volume = {18},
	number = {7},
	pages = {760--764},
	year = {2022},
	publisher = {Nature Publishing Group},
	doi = {10.1038/s41567-022-01619-7}
}

@article{Koh2023,
	author = {Koh, Jin Ming and Sun, Shi-Ning and Motta, Mario and Minnich, Austin J.},
	title = {{Measurement-induced entanglement phase transition on a superconducting quantum processor with mid-circuit readout}},
	journal = {Nat. Phys.},
	volume = {19},
	number = {9},
	pages = {1314--1319},
	year = {2023},
	publisher = {Nature Publishing Group},
	doi = {10.1038/s41567-023-02076-6}
}

@article{PhysRevLett.117.190503,
	author = {Sank, Daniel and Chen, Zijun and Khezri, Mostafa and Kelly, J. and Barends, R. and Campbell, B. and Chen, Y. and Chiaro, B. and Dunsworth, A. and Fowler, A. and others},
	title = {Measurement-Induced State Transitions in a Superconducting Qubit: Beyond the Rotating Wave Approximation},
	journal = {Phys. Rev. Lett.},
	volume = {117},
	number = {19},
	pages = {190503},
	year = {2016},
	publisher = {American Physical Society},
	doi = {10.1103/PhysRevLett.117.190503}
}

@article{PhysRevLett.126.170602,
  title = {Entanglement Transition in a Monitored Free-Fermion Chain: From Extended Criticality to Area Law},
  author = {Alberton, O. and Buchhold, M. and Diehl, S.},
  journal = {Phys. Rev. Lett.},
  volume = {126},
  issue = {17},
  pages = {170602},
  numpages = {6},
  year = {2021},
  month = {Apr},
  publisher = {American Physical Society},
  doi = {10.1103/PhysRevLett.126.170602},
  url = {https://link.aps.org/doi/10.1103/PhysRevLett.126.170602}
}

@article{PhysRevLett.115.140402,
  title = {Measurement-Induced Localization of an Ultracold Lattice Gas},
  author = {Patil, Y. S. and Chakram, S. and Vengalattore, M.},
  journal = {Phys. Rev. Lett.},
  volume = {115},
  issue = {14},
  pages = {140402},
  numpages = {5},
  year = {2015},
  month = {Oct},
  publisher = {American Physical Society},
  doi = {10.1103/PhysRevLett.115.140402},
  url = {https://link.aps.org/doi/10.1103/PhysRevLett.115.140402}
}

@article{PhysRevLett.97.260402,
  title = {Continuous and Pulsed Quantum {Z}eno Effect},
  author = {Streed, Erik W. and Mun, Jongchul and Boyd, Micah and Campbell, Gretchen K. and Medley, Patrick and Ketterle, Wolfgang and Pritchard, David E.},
  journal = {Phys. Rev. Lett.},
  volume = {97},
  issue = {26},
  pages = {260402},
  numpages = {4},
  year = {2006},
  month = {Dec},
  publisher = {American Physical Society},
  doi = {10.1103/PhysRevLett.97.260402},
  url = {https://link.aps.org/doi/10.1103/PhysRevLett.97.260402}
}

@article{PhysRevResearch.6.043246,
  title = {{Monitored fermions with conserved $U(1)$ charge}},
  author = {Fava, Michele and Piroli, Lorenzo and Bernard, Denis and Nahum, Adam},
  journal = {Phys. Rev. Res.},
  volume = {6},
  issue = {4},
  pages = {043246},
  numpages = {21},
  year = {2024},
  month = {Dec},
  publisher = {American Physical Society},
  doi = {10.1103/PhysRevResearch.6.043246},
  url = {https://link.aps.org/doi/10.1103/PhysRevResearch.6.043246}
}

@article{jppz-vdgn,
  title = {Generalized {Z}eno Effect and Entanglement Dynamics Induced by Fermion Counting},
  author = {Starchl, Elias and Fischer, Mark H. and Sieberer, Lukas M.},
  journal = {PRX Quantum},
  volume = {6},
  issue = {3},
  pages = {030302},
  numpages = {47},
  year = {2025},
  month = {Jul},
  publisher = {American Physical Society},
  doi = {10.1103/jppz-vdgn},
  url = {https://link.aps.org/doi/10.1103/jppz-vdgn}
}

@article{PhysRevA.86.032120,
  title = {Quantum {Z}eno dynamics of a field in a cavity},
  author = {Raimond, J. M. and Facchi, P. and Peaudecerf, B. and Pascazio, S. and Sayrin, C. and Dotsenko, I. and Gleyzes, S. and Brune, M. and Haroche, S.},
  journal = {Phys. Rev. A},
  volume = {86},
  issue = {3},
  pages = {032120},
  numpages = {15},
  year = {2012},
  month = {Sep},
  publisher = {American Physical Society},
  doi = {10.1103/PhysRevA.86.032120},
  url = {https://link.aps.org/doi/10.1103/PhysRevA.86.032120}
}

@article{Kondo2016,
	author = {Kondo, Yasushi and Matsuzaki, Yuichiro and Matsushima, Kei and Filgueiras, Jefferson G.},
	title = {{Using the quantum Zeno effect for suppression of decoherence}},
	journal = {New J. Phys.},
	volume = {18},
	number = {1},
	pages = {013033},
	year = {2016},
	publisher = {IOP Publishing},
	doi = {10.1088/1367-2630/18/1/013033}
}

@article{AubryAndre1980,
  title={Analyticity breaking and {A}nderson localization in incommensurate lattices},
  author={Aubry, Serge and Andr{\'e}, Gilles},
  journal={Ann. Israel Phys. Soc},
  volume={3},
  number={133},
  pages={18},
  year={1980}
}

@article{PhysRevLett.89.080401,
  title = {Quantum {Z}eno Subspaces},
  author = {Facchi, P. and Pascazio, S.},
  journal = {Phys. Rev. Lett.},
  volume = {89},
  issue = {8},
  pages = {080401},
  numpages = {4},
  year = {2002},
  month = {Aug},
  publisher = {American Physical Society},
  doi = {10.1103/PhysRevLett.89.080401},
  url = {https://link.aps.org/doi/10.1103/PhysRevLett.89.080401}
}

@article{PhysRevA.41.2295,
  title = {Quantum {Z}eno effect},
  author = {Itano, Wayne M. and Heinzen, D. J. and Bollinger, J. J. and Wineland, D. J.},
  journal = {Phys. Rev. A},
  volume = {41},
  issue = {5},
  pages = {2295--2300},
  numpages = {0},
  year = {1990},
  month = {Mar},
  publisher = {American Physical Society},
  doi = {10.1103/PhysRevA.41.2295},
  url = {https://link.aps.org/doi/10.1103/PhysRevA.41.2295}
}

@article{PhysRevResearch.2.033512,
  title = {Quantum {Z}eno effect appears in stages},
  author = {Snizhko, Kyrylo and Kumar, Parveen and Romito, Alessandro},
  journal = {Phys. Rev. Res.},
  volume = {2},
  issue = {3},
  pages = {033512},
  numpages = {12},
  year = {2020},
  month = {Sep},
  publisher = {American Physical Society},
  doi = {10.1103/PhysRevResearch.2.033512},
  url = {https://link.aps.org/doi/10.1103/PhysRevResearch.2.033512}
}

@article{Slichter_2016,
	author = {Slichter, D. H. and M{\ifmmode\ddot{u}\else\"{u}\fi}ller, C. and Vijay, R. and Weber, S. J. and Blais, A. and Siddiqi, I.},
	title = {{Quantum Zeno effect in the strong measurement regime of circuit quantum electrodynamics}},
	journal = {New J. Phys.},
	volume = {18},
	number = {5},
	pages = {053031},
	year = {2016},
	publisher = {IOP Publishing},
	doi = {10.1088/1367-2630/18/5/053031}
}

@article{Misra1977Zeno,
	author = {Misra, B. and Sudarshan, E. C. G.},
	title = {{The Zeno{'}s paradox in quantum theory}},
	journal = {J. Math. Phys.},
	volume = {18},
	number = {4},
	pages = {756--763},
	year = {1977},
	publisher = {AIP Publishing},
	doi = {10.1063/1.523304}
}

@article{g7vd-hgw4,
  title = {{Numerical study of the localization transition of Aubry-Andr\'e type models}},
  author = {Het\'enyi, Bal\'azs and Balogh, Istv\'an},
  journal = {Phys. Rev. B},
  volume = {112},
  issue = {14},
  pages = {144203},
  numpages = {12},
  year = {2025},
  month = {Oct},
  publisher = {American Physical Society},
  doi = {10.1103/g7vd-hgw4},
  url = {https://link.aps.org/doi/10.1103/g7vd-hgw4}
}

@article{3zfd-3hqt,
  title = {Measurement-induced phase transitions for free fermions in a quasiperiodic potential},
  author = {Matsubara, Toranosuke and Yamamoto, Kazuki and Koga, Akihisa},
  journal = {Phys. Rev. B},
  volume = {112},
  issue = {5},
  pages = {054309},
  numpages = {13},
  year = {2025},
  month = {Aug},
  publisher = {American Physical Society},
  doi = {10.1103/3zfd-3hqt},
  url = {https://link.aps.org/doi/10.1103/3zfd-3hqt}
}

@article{gisin1997quantumstatediffusionfoundations,
	author = {Gisin, Nicolas and Percival, Ian C.},
	title = {Quantum State Diffusion: from Foundations to Applications},
	journal = {arXiv},
	year = {1997},
	eprint = {quant-ph/9701024},
	doi = {10.48550/arXiv.quant-ph/9701024}
}

@article{Percival_1999,
	author = {Percival, Ian C.},
	title = {{Quantum state diffusion, measurement and second quantization}},
	journal = {Phys. Lett. A},
	volume = {261},
	number = {3},
	pages = {134--138},
	year = {1999},
	publisher = {North-Holland},
	doi = {10.1016/S0375-9601(99)00526-5}
}

@article{PhysRevA.69.032107,
  title = {Continuous measurement of canonical observables and limit stochastic {S}chr\"odinger equations},
  author = {Gough, John and Sobolev, Andrei},
  journal = {Phys. Rev. A},
  volume = {69},
  issue = {3},
  pages = {032107},
  numpages = {7},
  year = {2004},
  month = {Mar},
  publisher = {American Physical Society},
  doi = {10.1103/PhysRevA.69.032107},
  url = {https://link.aps.org/doi/10.1103/PhysRevA.69.032107}
}

@article{GisinPercival1992,
	author = {Gisin, N. and Percival, I. C.},
	title = {{The quantum state diffusion picture of physical processes}},
	journal = {J. Phys. A: Math. Gen.},
	volume = {26},
	number = {9},
	pages = {2245},
	year = {1993},
	publisher = {IOP Publishing},
	doi = {10.1088/0305-4470/26/9/019}
}

@book{Percival1998,
  title={Quantum State Diffusion},
  author={Percival, I. C.},
  year={1998},
  publisher={Cambridge University Press}
}

@article{PhysRevB.110.024303,
  title = {Unscrambling of single-particle wave functions in systems localized through disorder and monitoring},
  author = {Szyniszewski, Marcin},
  journal = {Phys. Rev. B},
  volume = {110},
  issue = {2},
  pages = {024303},
  numpages = {12},
  year = {2024},
  month = {Jul},
  publisher = {American Physical Society},
  doi = {10.1103/PhysRevB.110.024303},
  url = {https://link.aps.org/doi/10.1103/PhysRevB.110.024303}
}

@article{PhysRevB.102.094310,
  title = {Self-averaging in many-body quantum systems out of equilibrium: Approach to the localized phase},
  author = {Torres-Herrera, E. Jonathan and De Tomasi, Giuseppe and Schiulaz, Mauro and P\'erez-Bernal, Francisco and Santos, Lea F.},
  journal = {Phys. Rev. B},
  volume = {102},
  issue = {9},
  pages = {094310},
  numpages = {13},
  year = {2020},
  month = {Sep},
  publisher = {American Physical Society},
  doi = {10.1103/PhysRevB.102.094310},
  url = {https://link.aps.org/doi/10.1103/PhysRevB.102.094310}
}

@article{Gordillo_Guerrero_2007,
	author = {Gordillo-Guerrero, A. and Ruiz-Lorenzo, J. J.},
	title = {{Self-averaging in the three-dimensional site diluted Heisenberg model at the critical point}},
	journal = {J. Stat. Mech.: Theory Exp.},
	volume = {2007},
	number = {06},
	pages = {P06014},
	year = {2007},
	publisher = {IOP Publishing},
	doi = {10.1088/1742-5468/2007/06/P06014}
}

@article{PhysRevB.101.174312,
  title = {Self-averaging in many-body quantum systems out of equilibrium: Chaotic systems},
  author = {Schiulaz, Mauro and Torres-Herrera, E. Jonathan and P\'erez-Bernal, Francisco and Santos, Lea F.},
  journal = {Phys. Rev. B},
  volume = {101},
  issue = {17},
  pages = {174312},
  numpages = {16},
  year = {2020},
  month = {May},
  publisher = {American Physical Society},
  doi = {10.1103/PhysRevB.101.174312},
  url = {https://link.aps.org/doi/10.1103/PhysRevB.101.174312}
}

@article{PhysRevX.8.021005,
	author = {Touzard, S. and Grimm, A. and Leghtas, Z. and Mundhada, S. O. and Reinhold, P. and Axline, C. and Reagor, M. and Chou, K. and Blumoff, J. and Sliwa, K. M. and others},
	title = {Coherent Oscillations inside a Quantum Manifold Stabilized by Dissipation},
	journal = {Phys. Rev. X},
	volume = {8},
	number = {2},
	pages = {021005},
	year = {2018},
	publisher = {American Physical Society},
	doi = {10.1103/PhysRevX.8.021005}
}

@article{reis2021relationstratonovichitointegrals,
	author = {Goncalo dos Reis and Vadim Platonov},
	title = {{On the relation between Stratonovich and Ito integrals with functional integrands of conditional measure flows}},
	journal = {arXiv},
	year = {2021},
	eprint = {2111.03523},
	doi = {10.48550/arXiv.2111.03523}
}

@article{Pesce2013,
	author = {Pesce, Giuseppe and McDaniel, Austin and Hottovy, Scott and Wehr, Jan and Volpe, Giovanni},
	title = {{Stratonovich-to-It{\ifmmode\hat{o}\else\^{o}\fi} transition in noisy systems with multiplicative feedback}},
	journal = {Nat. Commun.},
	volume = {4},
	number = {2733},
	pages = {2733},
	year = {2013},
	publisher = {Nature Publishing Group},
	doi = {10.1038/ncomms3733}
}

@article{PhysRevB.110.184211,
  title = {Enhanced localization in the prethermal regime of continuously measured many-body localized systems},
  author = {Patrick, Kristian and Yang, Qinghong and Liu, Dong E.},
  journal = {Phys. Rev. B},
  volume = {110},
  issue = {18},
  pages = {184211},
  numpages = {10},
  year = {2024},
  month = {Nov},
  publisher = {American Physical Society},
  doi = {10.1103/PhysRevB.110.184211},
  url = {https://link.aps.org/doi/10.1103/PhysRevB.110.184211}
}

@article{PhysRevB.110.134203,
  title = {Asymmetric transfer matrix analysis of {L}yapunov exponents in one-dimensional nonreciprocal quasicrystals},
  author = {Li, Shan-Zhong and Cheng, Enhong and Zhu, Shi-Liang and Li, Zhi},
  journal = {Phys. Rev. B},
  volume = {110},
  issue = {13},
  pages = {134203},
  numpages = {11},
  year = {2024},
  month = {Oct},
  publisher = {American Physical Society},
  doi = {10.1103/PhysRevB.110.134203},
  url = {https://link.aps.org/doi/10.1103/PhysRevB.110.134203}
}

@article{PhysRevB.104.104203,
  title = {Transfer matrix study of the Anderson transition in non-Hermitian systems},
  author = {Luo, Xunlong and Ohtsuki, Tomi and Shindou, Ryuichi},
  journal = {Phys. Rev. B},
  volume = {104},
  issue = {10},
  pages = {104203},
  numpages = {20},
  year = {2021},
  month = {Sep},
  publisher = {American Physical Society},
  doi = {10.1103/PhysRevB.104.104203},
  url = {https://link.aps.org/doi/10.1103/PhysRevB.104.104203}
}

@article{Lunt2020,
	author = {Lunt, Oliver and Pal, Arijeet},
	title = {{Measurement-induced entanglement transitions in many-body localized systems}},
	journal = {Phys. Rev. Res.},
	volume = {2},
	number = {4},
	pages = {043072},
	year = {2020},
	publisher = {American Physical Society},
	doi = {10.1103/PhysRevResearch.2.043072}
}

@article{Boorman2022,
	author = {Boorman, T. and Szyniszewski, M. and Schomerus, H. and Romito, A.},
	title = {{Diagnostics of entanglement dynamics in noisy and disordered spin chains via the measurement-induced steady-state entanglement transition}},
	journal = {Phys. Rev. B},
	volume = {105},
	number = {14},
	pages = {144202},
	year = {2022},
	publisher = {American Physical Society},
	doi = {10.1103/PhysRevB.105.144202}
}

@article{Tang2025,
	author = {Tang, Yicheng and Kattel, Pradip and Pal, Arijeet and Yuzbashyan, Emil A. and Pixley, J. H.},
	title = {{The measurement-induced phase transition in strongly disordered spin chains}},
	journal = {arXiv},
	year = {2025},
	eprint = {2512.02100},
	doi = {10.48550/arXiv.2512.02100}
}

@article{Comtet2013,
	author = {Comtet, Alain and Texier, Christophe and Tourigny, Yves},
	title = {{Lyapunov exponents, one-dimensional Anderson localization and products of random matrices}},
	journal = {J. Phys. A: Math. Theor.},
	volume = {46},
	number = {25},
	pages = {254003},
	year = {2013},
	publisher = {IOP Publishing},
	doi = {10.1088/1751-8113/46/25/254003}
}

@book{Gardiner1985,
  author    = {C. W. Gardiner},
  title     = {Handbook of Stochastic Methods for Physics, Chemistry and the Natural Sciences},
  publisher = {Springer-Verlag},
  address   = {Berlin},
  year      = {1985}
}

@article{Poboiko2025,
	author = {Poboiko, Igor and Szyniszewski, Marcin and Turner, Christopher J. and Gornyi, Igor V. and Mirlin, Alexander D. and Pal, Arijeet},
	title = {{Measurement-induced L\'evy flights of quantum information}},
	journal = {Phys. Rev. Lett.},
	volume = {135},
	number = {17},
	pages = {170403},
	year = {2025},
	publisher = {American Physical Society},
	doi = {10.1103/tx71-1cd9}
}

@article{Turkeshi2021,
	author = {Turkeshi, Xhek and Biella, Alberto and Fazio, Rosario and Dalmonte, Marcello and Schir{\ifmmode\acute{o}\else\'{o}\fi}, Marco},
	title = {{Measurement-induced entanglement transitions in the quantum Ising chain: From infinite to zero clicks}},
	journal = {Phys. Rev. B},
	volume = {103},
	number = {22},
	pages = {224210},
	year = {2021},
	publisher = {American Physical Society},
	doi = {10.1103/PhysRevB.103.224210}
}

@article{Fidkowski2021,
	author = {Fidkowski, Lukasz and Haah, Jeongwan and Hastings, Matthew B.},
	title = {How Dynamical Quantum Memories Forget},
	journal = {Quantum},
	volume = {5},
	pages = {382},
	year = {2021},
	publisher = {Verein zur F{\ifmmode\ddot{o}\else\"{o}\fi}rderung des Open Access Publizierens in den Quantenwissenschaften},
	eprint = {2008.10611v2},
	doi = {10.22331/q-2021-01-17-382}
}

@article{Poboiko2024,
	author = {Poboiko, Igor and Gornyi, Igor V. and Mirlin, Alexander D.},
	title = {Measurement-Induced Phase Transition for Free Fermions above One Dimension},
	journal = {Phys. Rev. Lett.},
	volume = {132},
	number = {11},
	pages = {110403},
	year = {2024},
	publisher = {American Physical Society},
	doi = {10.1103/PhysRevLett.132.110403}
}

@article{Chan2019,
	author = {Chan, Amos and Nandkishore, Rahul M. and Pretko, Michael and Smith, Graeme},
	title = {{Unitary-projective entanglement dynamics}},
	journal = {Phys. Rev. B},
	volume = {99},
	number = {22},
	pages = {224307},
	year = {2019},
	publisher = {American Physical Society},
	doi = {10.1103/PhysRevB.99.224307}
}

@article{Anderson1958,
	author = {Anderson, P. W.},
	title = {Absence of Diffusion in Certain Random Lattices},
	journal = {Phys. Rev.},
	volume = {109},
	number = {5},
	pages = {1492--1505},
	year = {1958},
	publisher = {American Physical Society},
	doi = {10.1103/PhysRev.109.1492}
}

@article{Thouless1972,
	author = {Thouless, D. J.},
	title = {{A relation between the density of states and range of localization for one dimensional random systems}},
	journal = {J. Phys. C: Solid State Phys.},
	volume = {5},
	number = {1},
	pages = {77},
	year = {1972},
	publisher = {IOP Publishing},
	doi = {10.1088/0022-3719/5/1/010}
}

@article{Abrahams1979,
	author = {Abrahams, E. and Anderson, P. W. and Licciardello, D. C. and Ramakrishnan, T. V.},
	title = {Scaling Theory of Localization: Absence of Quantum Diffusion in Two Dimensions},
	journal = {Phys. Rev. Lett.},
	volume = {42},
	number = {10},
	pages = {673--676},
	year = {1979},
	publisher = {American Physical Society},
	doi = {10.1103/PhysRevLett.42.673}
}

@article{Basko2006,
	author = {Basko, D. M. and Aleiner, I. L. and Altshuler, B. L.},
	title = {{Metal{\textendash}insulator transition in a weakly interacting many-electron system with localized single-particle states}},
	journal = {Ann. Phys.},
	volume = {321},
	number = {5},
	pages = {1126--1205},
	year = {2006},
	publisher = {Academic Press},
	doi = {10.1016/j.aop.2005.11.014}
}

@article{Gornyi2005,
	author = {Gornyi, I. V. and Mirlin, A. D. and Polyakov, D. G.},
	title = {Interacting Electrons in Disordered Wires: {A}nderson Localization and Low-{$T$} Transport},
	journal = {Phys. Rev. Lett.},
	volume = {95},
	number = {20},
	pages = {206603},
	year = {2005},
	publisher = {American Physical Society},
	doi = {10.1103/PhysRevLett.95.206603}
}

@article{Pal2010,
	author = {Pal, Arijeet and Huse, David A.},
	title = {{Many-body localization phase transition}},
	journal = {Phys. Rev. B},
	volume = {82},
	number = {17},
	pages = {174411},
	year = {2010},
	publisher = {American Physical Society},
	doi = {10.1103/PhysRevB.82.174411}
}

@article{Nandkishore2015,
	author = {Nandkishore, Rahul and Huse, David A.},
	title = {Many-Body Localization and Thermalization in Quantum Statistical Mechanics},
	journal = {Annu. Rev. Condens. Matter Phys.},
	volume = {6},
	pages = {15--38},
	year = {2015},
	publisher = {Annual Reviews},
	doi = {10.1146/annurev-conmatphys-031214-014726}
}

@article{Abanin2019,
	author = {Abanin, Dmitry A. and Altman, Ehud and Bloch, Immanuel and Serbyn, Maksym},
	title = {{Colloquium: Many-body localization, thermalization, and entanglement}},
	journal = {Rev. Mod. Phys.},
	volume = {91},
	number = {2},
	pages = {021001},
	year = {2019},
	publisher = {American Physical Society},
	doi = {10.1103/RevModPhys.91.021001}
}

@article{Grempel1982,
	author = {Grempel, D. R. and Fishman, Shmuel and Prange, R. E.},
	title = {Localization in an Incommensurate Potential: An Exactly Solvable Model},
	journal = {Phys. Rev. Lett.},
	volume = {49},
	number = {11},
	pages = {833},
	year = {1982},
	publisher = {American Physical Society},
	doi = {10.1103/PhysRevLett.49.833}
}

@article{Thouless1983,
	author = {Thouless, D. J.},
	title = {{Bandwidths for a quasiperiodic tight-binding model}},
	journal = {Phys. Rev. B},
	volume = {28},
	number = {8},
	pages = {4272},
	year = {1983},
	publisher = {American Physical Society},
	doi = {10.1103/PhysRevB.28.4272}
}

@article{Iyer2013,
	author = {Iyer, Shankar and Oganesyan, Vadim and Refael, Gil and Huse, David A.},
	title = {{Many-body localization in a quasiperiodic system}},
	journal = {Phys. Rev. B},
	volume = {87},
	number = {13},
	pages = {134202},
	year = {2013},
	publisher = {American Physical Society},
	doi = {10.1103/PhysRevB.87.134202}
}

@article{Roati2008,
	author = {Roati, Giacomo and D{'}Errico, Chiara and Fallani, Leonardo and Fattori, Marco and Fort, Chiara and Zaccanti, Matteo and Modugno, Giovanni and Modugno, Michele and Inguscio, Massimo},
	title = {{Anderson localization of a non-interacting Bose{\textendash}Einstein condensate}},
	journal = {Nature},
	volume = {453},
	number = {7197},
	pages = {895--898},
	year = {2008},
	publisher = {Nature Publishing Group},
	doi = {10.1038/nature07071}
}

@article{Lahini2009,
	author = {Lahini, Y. and Pugatch, R. and Pozzi, F. and Sorel, M. and Morandotti, R. and Davidson, N. and Silberberg, Y.},
	title = {Observation of a Localization Transition in Quasiperiodic Photonic Lattices},
	journal = {Phys. Rev. Lett.},
	volume = {103},
	number = {1},
	pages = {013901},
	year = {2009},
	publisher = {American Physical Society},
	doi = {10.1103/PhysRevLett.103.013901}
}

@article{Deutsch1991,
	author = {Deutsch, J. M.},
	title = {{Quantum statistical mechanics in a closed system}},
	journal = {Phys. Rev. A},
	volume = {43},
	number = {4},
	pages = {2046--2049},
	year = {1991},
	publisher = {American Physical Society},
	doi = {10.1103/PhysRevA.43.2046}
}

@article{Srednicki1994,
	author = {Srednicki, Mark},
	title = {{Chaos and quantum thermalization}},
	journal = {Phys. Rev. E},
	volume = {50},
	number = {2},
	pages = {888--901},
	year = {1994},
	publisher = {American Physical Society},
	doi = {10.1103/PhysRevE.50.888}
}

@article{DAlessio2016,
	author = {D'Alessio, Luca and Kafri, Yariv and Polkovnikov, Anatoli and Rigol, Marcos},
	title = {{From quantum chaos and eigenstate thermalization to statistical mechanics and thermodynamics}},
	journal = {Adv. Phys.},
    volume = {65},
    number = {3},
    pages = {239--362},
	year = {2016},
	publisher = {Taylor {\&} Francis},
    doi = {10.1080/00018732.2016.1198134},
	url = {https://doi.org/10.1080/00018732.2016.1198134}
}

@article{Borgonovi2016,
	author = {Borgonovi, F. and Izrailev, F. M. and Santos, L. F. and Zelevinsky, V. G.},
	title = {{Quantum chaos and thermalization in isolated systems of interacting particles}},
	journal = {Phys. Rep.},
	volume = {626},
	pages = {1--58},
	year = {2016},
	publisher = {North-Holland},
	doi = {10.1016/j.physrep.2016.02.005}
}

@article{DeTomasi2024,
	author = {De Tomasi, Giuseppe and Khaymovich, Ivan M.},
	title = {{Stable many-body localization under random continuous measurements in the no-click limit}},
	journal = {Phys. Rev. B},
	volume = {109},
	number = {17},
	pages = {174205},
	year = {2024},
	publisher = {American Physical Society},
	doi = {10.1103/PhysRevB.109.174205}
}

@article{CombesThomas1973,
	author = {Combes, J. M. and Thomas, L.},
	title = {{Asymptotic behaviour of eigenfunctions for multiparticle Schr{\ifmmode\ddot{o}\else\"{o}\fi}dinger operators}},
	journal = {Commun. Math. Phys.},
	volume = {34},
	number = {4},
	pages = {251--270},
	year = {1973},
	publisher = {Springer-Verlag},
	doi = {10.1007/BF01646473}
}

@dataset{Data,
    author = {Singha, Pinaki and Roy, Nilanjan and Szyniszewski, Marcin and Sharma, Auditya},
    title = {{Research data for ``Controlled Zeno-Induced Localization of Free Fermions in a Quasiperiodic Chain''}},
    year = {2026},
    url = {https://doi.org/10.25446/oxford.32172429},
    doi = {10.25446/oxford.32172429}
}

\end{document}